\documentclass[12pt]{article}

\usepackage{amsmath}
\usepackage{amssymb}
\usepackage{amsfonts}
\usepackage{mathrsfs}
\usepackage{graphicx}
\usepackage{enumerate}

\newcommand{\ket}[1]{|#1\rangle}

\usepackage[margin=1in]{geometry}
\usepackage[width=0.9\textwidth]{caption}

\usepackage[pdftex,colorlinks=true,linkcolor=blue,citecolor=blue,urlcolor=blue]{hyperref}

\begin{document}

\title{Complexity in de Sitter space}
\author{Alan Reynolds\footnote{a.p.reynolds@durham.ac.uk}\,\,  and Simon F. Ross\footnote{s.f.ross@durham.ac.uk} \\  \bigskip \\ Centre for Particle Theory, Department of Mathematical Sciences \\ Durham University\\ South Road, Durham DH1 3LE}

\maketitle
 
\begin{abstract}
We consider the holographic complexity conjectures for de-Sitter invariant states in a quantum field theory on de Sitter space, dual to asymptotically anti-de Sitter geometries with de Sitter boundaries. The bulk holographic duals include solutions with or without a horizon. If we compute the complexity from the spatial volume, we find results consistent with general expectations, but the conjectured bound on the growth rate is not saturated. If we compute complexity from the action of the Wheeler-de Witt patch, we find qualitative differences from the volume calculation, with states of smaller energy having larger complexity than those of larger energy, even though the latter have bulk horizons.
\end{abstract}
 
\clearpage 

\section{Introduction}
 
 In principle, holography provides a well-defined non-perturbative formulation of quantum gravity. But to really use it to address questions about the nature of spacetime, we need to understand the emergence of the bulk spacetime from the dual field theory description. Since the conjecture of Ryu and Takayanagi \cite{Ryu:2006bv}, there has been growing evidence that entanglement plays an important role, and a variety of tools from quantum information have been applied to understand how spacetime emerges from the field theory. In \cite{Susskind:2014rva}, Susskind conjectured a new relation between the bulk geometry and the dual boundary state, proposing that the time-dependent geometry of the region behind the horizon of an AdS black hole could be related to the complexity of the dual boundary state.\footnote{The quantum computational complexity is a measure of the minumum number of elementary gates needed in a quantum circuit which constructs a given state  starting from a specified simple reference state (see e.g. \cite{Osborne:c}).}
This proposal was refined  in \cite{Stanford:2014jda} into the conjecture that the computational complexity of the boundary state at a given time (on some spacelike slice of the boundary) could be identified with the volume of a maximal volume spacelike slice in the bulk, ending on the given boundary slice. This will be referred to as the CV conjecture. This was further developed in \cite{Susskind:2014jwa,Susskind:2014moa}. 

More recently, it was conjectured that the complexity is related instead to the action of a Wheeler-de Witt patch in the bulk bounded by the given spacelike surface \cite{Brown:2015bva,Brown:2015lvg}. This is referred to as the CA conjecture. An appropriate prescription for calculating the action for a region of spacetime bounded by null surfaces was obtained in \cite{Lehner:2016vdi}. Further related work is \cite{Barbon:2015ria,Brown:2016wib,Couch:2016exn,Yang:2016awy,Chapman:2016hwi}.
 
No derivation of these conjectures, relating them back to the basic AdS/CFT dictionary, has yet been given. They are supported by the relation between the results of the bulk calculation and general expectations for the behaviour of the complexity in a generic quantum system. This evidence comes so far from the study of black hole spacetimes. Both the CV and CA conjectures produce results for the complexity that grow linearly in time, coming from the contributions from the region behind the black hole horizon. This linear growth is supposed to be generic for interacting quantum systems in states of non-maximal complexity \cite{Brown:2015lvg,Brown:2017jil}. Furthermore, the time derivative is simply proportional to the mass of the black hole, which can be interpreted as the energy of the state in the dual theory. This saturates a proposed bound on the growth of the complexity \cite{lloyd,Brown:2015lvg}. From the bulk point of view, it is highly non-trivial that one obtains simply the mass. This relation to the mass persists in studies of the effects of higher-curvature corrections \cite{Tao:2017fsy,Pan:2016ecg,Alishahiha:2017hwg,Wang:2017uiw,Guo:2017rul}, although recent work on the inclusion of flavour branes finds that the bound is satisfied but no longer saturated \cite{Abad:2017cgl}.
 
It is clearly of interest to consider the conjecture in other settings. In this paper, we consider solutions which are asymptotically AdS in a de Sitter slicing, dual to field theories in de Sitter space. This is an appealing test case firstly because it provides a simple example where the field theory is in a time-dependent background, so we can study how the time-dependence of the complexity of the state is affected by time-dependent sources, and secondly because simple bulk duals are known.

This situation also has some features in common with the black hole case.  In the field theory, we  consider some de Sitter-invariant state on a de Sitter background. This is a globally pure state, but the state seen by a given observer will appear thermal. This thermal structure comes from entanglement between degrees of freedom at antipodal points in the de Sitter space, whose structure is similar to that between the two copies in a thermofield double state. In the bulk, the geometries dual to some such states have horizons, while other cases do not, so they provide another example where one can study the relation of bulk horizons and the behind the horizon geometry to the complexity. The entanglement structure in this case was investigated in \cite{Maldacena:2012xp}, which provided important inspiration for our work. Complexity of these solutions was also previously considered in \cite{Barbon:2015ria}. 

In section \ref{review}, we review the CV and CA conjectures, and their application to the black hole examples. In section \ref{desitter}, we discuss the general features we would expect in the complexity in de Sitter space from the field theory point of view, and set up the holographic calculation, discussing general features of the asymptotically AdS spaces we consider and introducing a particular set of $\text{de Sitter} \times S^1$ examples we will focus on. We find that if we work in flat coordinates on the de Sitter space, the time dependence of the complexity is fixed by the de Sitter symmetry. This is a surprisingly simple behaviour, and qualitatively different from the time dependence of the entanglement entropy found in \cite{Maldacena:2012xp}. For the $\text{de Sitter} \times S^1$ examples, there are explicitly known bulk geometries, which either end in a bubble of nothing, or have a bulk horizon, with a crunching FRW region beyond it. 

In section \ref{volume}, we discuss the  calculation using the CV conjecture. We find that the growth rate for the complexity for the bubbles is smaller than for the black holes, as the spatial slice ends on the bubble. This is consistent with the proposed relation of growth rate to energy, as the bubbles have smaller energy than the black hole, although the bound is not saturated; for large bubbles the difference in complexity grows more slowly than the difference in energy.  

In section \ref{action}, we perform  calculations using the CA conjecture. We find that the growth rate for the complexity for the bubbles in this case is {\it larger} than for the black holes: this is due to a negative spacetime volume contribution to the action. This is contrary to the general expectations for the behaviour of the complexity. Furthermore, for small bubbles, there is a logarithmic growth in the complexity as the bubble shrinks, whereas the volume calculation approaches a finite limit. We note that the prescription we use for calculating the action is not unique, and consider options for modifying it. We give concluding remarks in section \ref{disc}. 

\section{Review of CV and CA}
\label{review}

We first review the two proposals for the holographic calculation of the complexity. We will discuss the CV proposal  briefly, focusing on the issue of UV divergences, and then consider the CA proposal in some detail, spelling out the prescription for the calculation of the action we will employ. 

In the CV conjecture of \cite{Susskind:2014rva}, the complexity $\mathcal C$ of a pure state $|\Psi \rangle$ of a holographic field theory on some spatial slice $\Sigma$ on the boundary of an asymptotically AdS spacetime is identified with the volume $V$ of the maximal volume codimension one slice $B$ in the bulk having its boundary on $\Sigma$,
\begin{equation}
{\mathcal C_{\textnormal{V}}} \propto \frac{V(B)}{G_{\textnormal{N}} l_{\textnormal{AdS}}}.
\end{equation}
%
This was motivated by the study of the behaviour of Schwarzschild-AdS black hole solutions, where it was found that the volume of the maximal volume slice grows linearly with time, even at late boundary times when other observables have thermalized. The complexity conjecture relates this linear growth to linear growth of the complexity of the dual state, which is expected to continue for exponentially long times in a generic interacting theory, starting from a low-complexity initial state (see \cite{Brown:2017jil} for a recent discussion of this growth of complexity).  
 
 The volume of the maximal volume slice has a divergence proportional to the volume of $\Sigma$. The work on black holes focused on the rate of change of the complexity, which is not sensitive to UV divergences, but in our context it will be important to understand these divergences, as the volume of $\Sigma$ will be time-dependent, so one cannot simply eliminate the divergence by considering the rate of change of complexity. One could remove the divergences in the bulk calculation by adding appropriate counterterms, as in holographic renormalization \cite{Kim:2017lrw}, or by subtracting some appropriate background, defining a ``complexity of formation'' \cite{Chapman:2016hwi}, but we would prefer to argue that these divergences should be identified as part of the physical complexity of the field theory state, just as the divergences in the area of bulk minimal surfaces can be identified with physical UV contribution in the entanglement entropy \cite{Ryu:2006bv}. 
 
We therefore want to argue that the divergence in the holographic calculation has a physical interpretation, as reflecting the difference in short-distance structure between the  state $|\Psi \rangle$ and the chosen reference state. A simple argument that produces a volume divergence starts by modelling the field theory as a lattice and taking the reference state to be a simple product state on the lattice sites. Any Hadamard state in the field theory will differ from this product state due to the short-range entanglement and correlation implied by the absence of arbitrarily high energy excitations. Setting up this entangled state from the reference product state would require some finite number of elementary operations per lattice site, producing a contribution to the complexity proportional to the volume of the field theory in units of the UV cutoff. The holographic calculation will in general contain subleading divergences, related to the curvature of $\Sigma$, as calculated for example in \cite{Carmi:2016wjl}. It would be interesting to explore the field theory interpretation of these subleading divergences, and whether they can be related to the structure of a generic Hadamard state. 
 
In \cite{Brown:2015bva,Brown:2015lvg}, an alternative CA conjecture was proposed. This identifies the complexity of $|\Psi \rangle$ with the action of the ``Wheeler-de Witt patch", the domain of development of the slice $B$ considered previously. The proposal is that
\begin{equation} \label{ca}
{\mathcal C_{\textnormal{A}}} = \frac{S_{\textnormal{W}}}{\pi \hbar},
\end{equation}
where $S_W$ is the action of the Wheeler-de Witt patch. This proposal has the advantage that the formula is more universal, containing no explicit reference to a bulk length scale. It is also often easier to calculate, as there is no maximisation problem to solve. Finding the Wheeler-de Witt patch for a given boundary slice is easier than finding the maximal volume slice. 

For the black hole solutions, the action of the Wheeler-de Witt patch turns out to also exhibit linear growth in time. In \cite{Brown:2015bva,Brown:2015lvg}, it was argued that the black hole saturates a conjectured  universal upper bound on the rate of growth of the complexity \cite{lloyd}
 \begin{equation} \label{ctd}
 \frac{d \mathcal C}{dt} \leq \frac{2 M}{\pi \hbar}. 
 \end{equation}
From the field theory point of view, the mass $M$ is the energy of the state. The saturation says that black holes represent the situation where the complexity is growing at its maximal rate, analogous to the conjecture that black holes are the fastest scramblers in nature \cite{Hayden:2007cs,Sekino:2008he,Maldacena:2015waa}. Further work on the CA proposal for charged black holes is found in \cite{Cai:2016xho,Cai:2017sjv,Tao:2017fsy}, while extensions to black holes in more general gravitational theories are found in \cite{Pan:2016ecg,Alishahiha:2017hwg,Wang:2017uiw,Guo:2017rul}. 

This bound is saturated by all Schwarzschild-AdS black holes in the CA conjecture \cite{Brown:2015bva,Brown:2015lvg}. It is also saturated in the CV conjecture for large Schwarzschild-AdS black holes if we take an appropriate normalization of the complexity in the latter case, 
\begin{equation} \label{cvnorm}
\mathcal C_{\textnormal{V}} = \frac{(d-1) V}{2 \pi^2 G_{\textnormal{N}} \ell} = \frac{8 (d-1) V}{\pi \ell}, 
\end{equation}
where in the second equality we adopt units where $16 \pi G_{\textnormal{N}} = 1$, as we will do henceforth. We will adopt this normalization of the complexity in the CV conjecture for definiteness in our later calculations. 

To apply the CA conjecture, we need to define a prescription for the calculation of the action. The Wheeler-de Witt patch has null boundaries, for which the appropriate boundary terms needed for the Einstein-Hilbert action were not yet known. In \cite{Lehner:2016vdi}, inspired by the CA conjecture, a prescription for the action of a region of spacetime containing null boundaries was constructed (see also \cite{Parattu:2015gga,Jubb:2016qzt}). The prescription was obtained by requiring that the variation of the action vanish on-shell when the variation of the metric vanishes on the boundary of the region. The resulting form for the action is (recalling that we work in units where $16 \pi G_{\textnormal{N}}=1$)  
\begin{eqnarray} \label{watact}
S_{\mathcal V} &=& \int_{\mathcal V} (R- 2 \Lambda) \sqrt{-g}\, dV + 2 \sum_{T_i} \int_{T_i} K\, d\Sigma + 2 \sum_{S_i} \mathrm{sign}(S_i) \int_{S_i}\, K d \Sigma \\ \nonumber
&& - 2 \sum_{N_i} \mathrm{sign}(N_i) \int_{N_i} \kappa\, dS \,d\lambda + 2 \sum_{j_i} \mathrm{sign}(j_i) \oint \eta_{j_i}\, dS + 2 \sum_{m_i} \mathrm{sign}(m_i) \oint a_{m_i}\, dS.
\end{eqnarray}
In this expression
\begin{itemize}
\item $T_i$ and $S_i$ are, respectively, timelike and spacelike components of the boundary of the region $\mathcal V$, and $K$ is the trace of the extrinsic curvature of the boundary. For $T_i$, the normal is taken outward-directed from $\mathcal V$. For $S_i$, the normal is always taken future-directed, and sign$(S_i) = 1(-1)$ if $\mathcal V$ lies to the future (past) of $S_i$, that is if the normal vector points into (out of) the region of interest. 
\item $N_i$ are null components of the boundary of $\mathcal V$, $\lambda$ is a parameter on null generators of $N_i$, increasing to the future, $dS$ is an area element on the cross-sections of constant $\lambda$, and $k^\alpha \nabla_\alpha k^\beta = \kappa k^\beta$, where $k^\alpha= \partial x^\alpha/\partial \lambda$ is the tangent to the generators. sign$(N_i) = 1(-1)$ if $N_i$ lies to the future (past) of   ${\mathcal V}$. 
\item $j_i$ are junctions between non-null boundary components,  where $\eta$ is the logarithm of the dot product of normals. We do not give the rules in detail as such junctions do not occur for Wheeler-de Witt patches in simple spacetimes; see \cite{Lehner:2016vdi} for full detail.
\item $m_i$ are junctions where one or both of the boundary components are null. We have a null surface with future-directed tangent $k^\alpha$ and either a spacelike surface with future directed unit normal $n^\alpha$, a timelike surface with outward directed unit normal $s^\alpha$, or another null surface with future-directed tangent $\bar k^\alpha$, and
\begin{equation}
a = \left\{ \begin{array}{c} \ln | k \cdot n| \\ \ln| k \cdot s| \\ \ln|k \cdot \bar k/2| \end{array} \right.
\end{equation}
respectively. sign$(m_i) = +1$ if $\mathcal V$ lies to the future (past) of the null boundary component and $m_i$ is at the past (future) end of the null component, and sign$(m_i) = -1$ otherwise.
\end{itemize}

An additional term was identified in  \cite{Lehner:2016vdi} which is not required by the variational argument, but which eliminates dependence on the parameter along the generators of the null surfaces, 
\begin{equation} \label{acorr}
\Delta S = -2 \sum_{N_i}  \mathrm{sign}(N_i) \int_{N_i} \Theta \ln |\ell \Theta|\mathop{dS}\mathop{d\lambda}, 
\end{equation}
where $\Theta$ is the expansion of the null generators of $N_i$, 
\begin{equation}
\Theta = \frac{1}{\sqrt{\gamma}} \frac{\partial \sqrt{\gamma}}{\partial \lambda},
\end{equation}
where $\gamma$ is the metric on the cross-sections of constant $\lambda$. In previous work \cite{Reynolds:2016rvl}, we showed that including this term has the additional virtue that it removes the leading divergences in the action \eqref{watact}. We will henceforth adopt the action $S = S_{\mathcal V} + \Delta S$ as our definition of the action for a region with null boundaries. This gives a complexity with a divergence proportional to the volume of the boundary slice, as in the CV case earlier. Note, however, that the structure of subleading divergences in the two cases is distinct \cite{Reynolds:2016rvl,Kim:2017lrw}. 

We will find later that this prescription for the action does not produce satisfactory results for solutions with de Sitter boundaries, so it is important to note that it is not unique. The prescription of \cite{Lehner:2016vdi}, demanding that the action have vanishing variation on-shell under variations of the metric which leave the intrinsic geometry of the boundary fixed, allows the addition of arbitrary boundary terms which are functions only of the intrinsic geometry of the boundary. On the null boundaries, this includes reparametrization-invariant terms of the form 
\begin{equation} 
S_{\textnormal{N}} =  \int_{N_i} \Theta f(\gamma)\mathop{dS}\mathop{d\lambda}, 
\end{equation}
where $f(\gamma)$ is any scalar function of the intrinsic metric $\gamma$ on the cross-sections of constant $\lambda$ and its derivatives; for example, the Ricci scalar of $\gamma$. We will not consider adding such terms here, but considering their impact is an important direction for future work, as we discuss later. 

The simplest example of the calculation, which will also be useful later, is to consider vacuum AdS$_{d+1}$ in Poincar\'e coordinates,
\begin{equation} \label{adsp}
ds^2 = \frac{\ell^2}{z^2} ( dz^2 - dt^2 + d\vec{x}^2),
\end{equation}
which is dual to the field theory in flat space. We consider a $d+1$ dimensional AdS space, with a $d$ dimensional boundary. 

For the CV conjecture, the maximal volume slice with boundary at $t=0$ is simply the $t=0$ surface in the bulk, whose volume is 
\begin{equation} \label{vol}
V(B) = \int \mathop{dz}\mathop{d^{d-1}x} \sqrt{h} = \ell^d V_x  \int_\epsilon^\infty \frac{\mathop{dz}}{z^d} = \frac{\ell^d V_x}{(d-1) \epsilon^{d-1}},
\end{equation}
where $V_x$ is the IR divergent coordinate volume in the $\vec x$ directions. Thus, the complexity calculated according to the CV prescription is, with the normalization of \eqref{cvnorm},
\begin{equation}  \label{cvol}
\mathcal{C}_{\textnormal{V}} =  \frac{8 \ell^{d-1} V_x}{\pi \epsilon^{d-1}}.
\end{equation}
This is proportional to the volume of the space the field theory lives in, in units of the cutoff. This has both an IR and a UV divergence, which is physically reasonable if we think of the complexity as defined with respect to some product lattice state, as previously discussed. 

\begin{figure}[htb]
\centering
\includegraphics[width=0.4\textwidth]{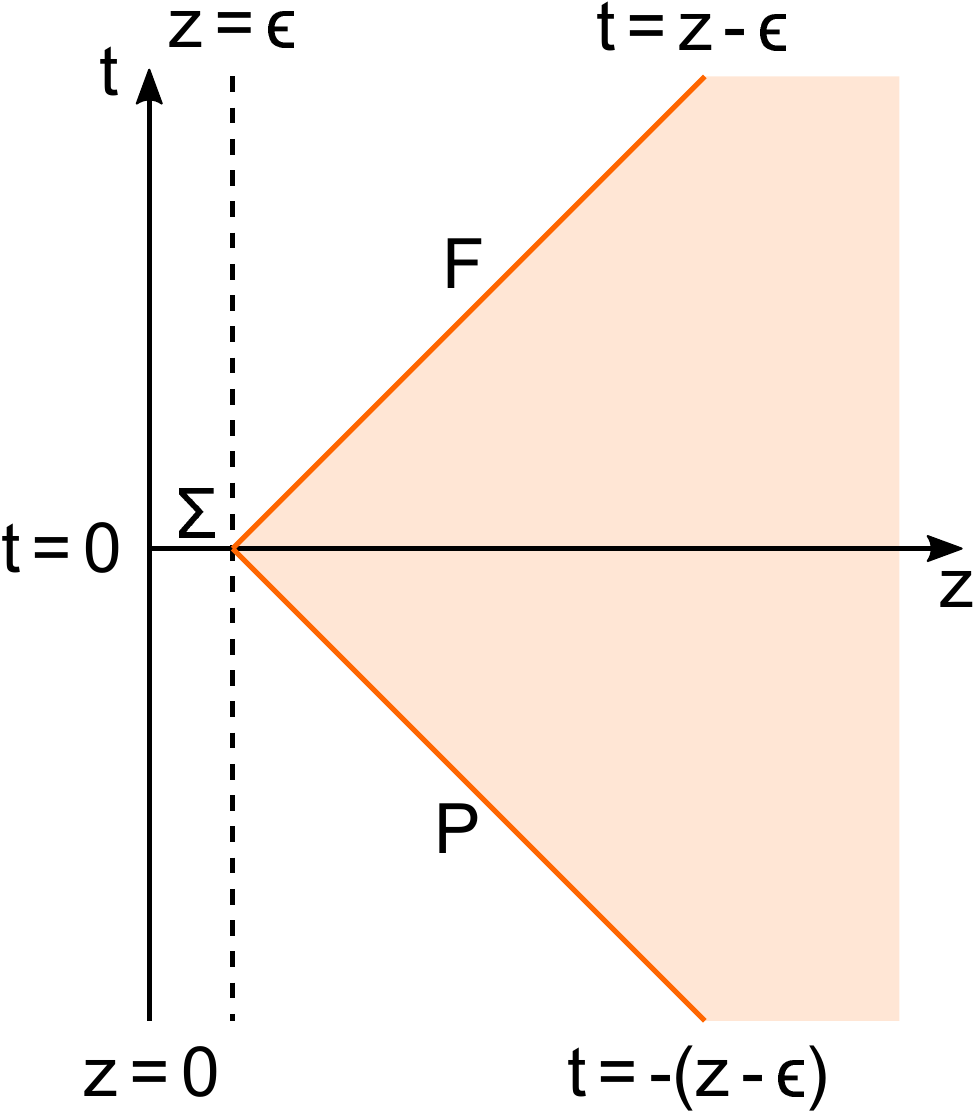}
\caption{The Wheeler-de Witt patch in Poincar\'e coordinates, showing future and past boundaries $F$ and $P$ and the surface $\Sigma$ at $t = 0, z = \epsilon$.}
\label{WdW in Poincare}
\end{figure}

Turning to the CA conjecture, consider the Wheeler-de Witt patch of this cutoff surface. If we ask for the complexity of the field theory on the $t=0$ surface, cut off at $z = \epsilon$, the Wheeler de Witt patch lies between $t = z-\epsilon$ and $t = -(z-\epsilon)$, as shown in figure \ref{WdW in Poincare}.
Note that although these coordinates do not cover the full spacetime, the Wheeler-de Witt patch lies inside the region covered by this coordinate patch. The action of the Wheeler-de Witt patch with the prescription of \cite{Lehner:2016vdi} is 
\begin{equation} \label{adsa}
S_{\cal V} = \int_{W} (R- 2 \Lambda) \sqrt{-g} \mathop{dV}  - 2  \int_{F} \kappa \mathop{dS}\mathop{d\lambda}   + 2  \int_{P} \kappa \mathop{dS}\mathop{d\lambda}  -  2   \oint_{\Sigma} a\mathop{dS}, 
\end{equation}
where $F$ is the future null boundary of the Wheeler-de Witt patch, $P$ is the past boundary and $\Sigma$ is the surface at $t=0$, $z=\epsilon$. Since $R - 2 \Lambda = -2d/\ell^2$, the volume integral is 
\begin{equation}
S_{\textnormal{Vol}} = - 2 \frac{d}{\ell^2} \int_\epsilon^\infty \mathop{dz} \int_{-(z-\epsilon)}^{z- \epsilon} \mathop{dt} \frac{\ell^{d+1}}{z^{d+1}} V_x = -4 \frac{\ell^{d-1} V_x}{(d-1) \epsilon^{d-1}}.
\end{equation}
In calculating the surface contributions, it is convenient to adopt an affine parametrization of the null surfaces, so that the integrals over the future and past boundaries do not contribute. Let us take the affine parameters along the null surfaces to be 
\begin{equation} \label{afp}
\lambda = - \frac{\ell^2}{\alpha z} \mbox{ on }F, \quad \lambda = \frac{\ell^2}{\beta z} \mbox{ on }P, 
\end{equation}
where we introduce the arbitrary constants $\alpha, \beta$ to exhibit explicitly the remaining coordinate dependence. This gives $k = \frac{\alpha z^2}{\ell^2} (\partial_t + \partial_z)$, $\bar k = \frac{\beta z^2}{\ell^2} (\partial_t - \partial_z)$.  The boundary corner term is thus
\begin{equation}
S_{\Sigma} = - 2 \frac{\ell^{d-1} V_x}{\epsilon^{d-1}} \ln(\alpha \beta \epsilon^2/\ell^2).  
\end{equation}
Thus, the action calculated according to \eqref{adsa} is 
\begin{equation}
S_{\cal V} = \frac{\ell^{d-1} V_x}{\epsilon^{d-2}}[ - 4 \ln( \epsilon/\ell) - 2 \ln(\alpha\beta) - \frac{1}{d-1} ].  
\end{equation}
We include the additional contribution \eqref{acorr}. The metric on $F$ has $\sqrt{\gamma} = \ell^{d-1}/z^{d-1}$, so the expansion is
\begin{equation}
\Theta = \frac{1}{\sqrt{\gamma}} \frac{\partial \sqrt{\gamma}}{\partial \lambda} = - \frac{1}{\sqrt{\gamma}} \alpha  \frac{z^2}{\ell^2}  \frac{\partial \sqrt{\gamma}}{\partial z} = (d-1) \alpha \frac{z}{\ell^2}, 
\end{equation}
so the surface term is
\begin{eqnarray}
S_{\textnormal{F}} &=& -2 (d-1) \ell^{d-1} V_x \int z^{-(d-2)} \ln (\alpha (d-1) z/\ell) \alpha \mathop{d\lambda} \\ &=&  2 (d-1) \ell^{d-1} V_x \int_\epsilon^\infty z^{-d} \ln (\alpha (d-1) z/\ell) \mathop{dz} \nonumber \\  &=&  2\frac{ \ell^{d-1}}{\epsilon^{d-1}} V_x \left( \ln( \alpha (d-1) \epsilon/\ell) + \frac{1}{d-1} \right),  \nonumber 
\end{eqnarray}
and similarly $S_{\textnormal{P}} = 2\frac{ \ell^{d-1}}{\epsilon^{d-1}} V_x \left( \ln(\beta (d-1) \epsilon/\ell) + \frac{1}{d-1} \right)$, so 
\begin{equation} \label{adsaf}
S = S_{\cal V} + \Delta S = S_{\textnormal{Vol}} + S_\Sigma + S_{\textnormal{F}} + S_{\textnormal{P}} = 4  \frac{ \ell^{d-1}}{\epsilon^{d-1}} V_x \ln(d-1).  
\end{equation}
Taking the proposal \eqref{ca} for the complexity, this gives
\begin{equation} 
{\mathcal C_{\textnormal{A}}} = \frac{ 4 \ell^{d-1} V_x}{\pi \epsilon^{d-1}}  \ln(d-1).  
\end{equation}
This has the same divergence structure as in \eqref{cvol}, but with a different coefficient. Subleading divergences in both calculations can be expressed in terms of local geometric properties of the boundary in both cases, but the relative coefficients differ.

\section{de Sitter complexity}
\label{desitter}

In this section we set up the calculation of complexity for field theories in de Sitter. We first consider the expectations from the field theory side, and then describe the bulk geometries dual to field theories in de Sitter, which will be used in the following sections to evaluate the complexity using the CV and CA conjectures.  

We will find it convenient to analyse the complexity in conformally flat coordinates on the de Sitter space, 
\begin{equation} \label{cfcoord}
ds^2 = \frac{1}{H^2 \eta^2} (- d\eta^2 + d\vec{x}^2),
\end{equation}
as there is a symmetry relating surfaces of different $\eta$. The surfaces of $\eta = $ constant are Cauchy surfaces for de Sitter, so we are measuring the complexity of the global pure state on the full de Sitter space. This is however a different physical question from asking about the $\tau = $ constant slices in global coordinates, 
\begin{equation}
ds^2 = H^{-2} (-d\tau^2 + \cosh^2 \tau \, d\Omega_{d-2}^2),
\end{equation}
as they are different spacelike surfaces in the de Sitter space. It would be interesting to also study the situation in global coordinates, but we will not explore this here.  We will consider the flat patch including the future boundary of de Sitter space, so in \eqref{cfcoord}, $\eta \in (-\infty, 0)$, with the future conformal boundary at $\eta=0$. We will write formulas for a $d-1$ dimensional de Sitter space, as in our most prominent examples, we take the $d$-dimensional boundary to be de Sitter$_{d-1} \times S^1$. 

We have argued above that a volume law divergence for complexity in field theory is a generic expectation. In the context of de Sitter, this contribution to the complexity will grow with time, as the proper volume of the spatial slices of the universe grows. Relative to a fixed cutoff, the number of lattice sites will grow with the volume, and we would expect the UV divergent part of the complexity to be proportional to the proper volume.  In the flat coordinates \eqref{cfcoord}, this proper volume is infinite, due to the infinite volume in the spatial $\vec x$ directions. This is an IR divergence in addition to the UV divergence, as in the Poincar\'e-AdS calculation in the previous section. 

In the holographic calculations, the volume of the bulk surface or action of the Wheeler-de Witt patch will involve an integral over these spatial directions, so the IR divergence is simply the volume $V_x$ in these directions, multiplied by some overall factor. It is plausible that this is true more generally, i.e. that $\mathcal C \propto V_x$. Whenever this holds, we can use this simple relation to fix the time dependence of the complexity by symmetry. In the flat coordinates \eqref{cfcoord}, there is a symmetry under $\eta \to \lambda \eta$, $\vec x \to \lambda \vec x$. The complexity of a de-Sitter invariant state should be invariant under this symmetry, which implies that if it is proportional to the spatial volume, the complexity of the state measured on a slice of fixed $\eta$ must be 
\begin{equation} \label{csymm}
\mathcal C = \frac{V_x}{|\eta|^{d-2}} c(\ket{\Psi}),
\end{equation}
where $c(\ket{\Psi})$ is independent of time. 

More generally, the holographic results for homogeneous boundary spaces will always be of the form 
\begin{equation} \label{cdens}
\mathcal C = V_\Sigma c
\end{equation}
where $V_\Sigma$ is the proper volume of the boundary spatial slice, and $c$ is a ``complexity density''.  In the black hole case this was a non-trivial function of time, which becomes linear at late times. For a de Sitter-invariant state the complexity density is constant by virtue of the symmetries. 

This behaviour is qualitatively different to that seen in the study of the entanglement entropy in de Sitter in \cite{Maldacena:2012xp}. There, they found that the entanglement entropy of a finite region in the $\vec x$ coordinates in the spatial slice at constant $\eta$ generically has a finite part with time dependence 
\begin{equation}
S_{\textnormal{UV} - \textnormal{finite}} = c_5 \frac{A_x}{\eta^{d-3}} + c_6 \log \eta + (\mbox{indep of }\eta),
\end{equation}
where $A_x$ is the coordinate area, in the spatial $\vec x$ coordinates, of the boundary of the region considered. The first term is analogous to our result, while the second provided a signal of the existence of horizons in the bulk. The complexity is simpler because the holographic calculations always give us a complexity simply proportional to the volume $V_x$. The complexity associated with subregions, as considered in \cite{Alishahiha:2015rta,Ben-Ami:2016qex,Bakhshaei:2017qud}, could have a more complicated behaviour, but since we have less understanding of the expectations of the behaviour of subregion complexity, we will not consider it here. 

This simplicity of the time dependence does not mean that everything is fixed by symmetry. As we will review below, in the holographic context there are multiple solutions with the same de Sitter asymptotics, corresponding to different de Sitter-invariant states on the same background. We would expect the difference between these states to be reflected in the complexity. Prompted by the bound \eqref{ctd}, we will compare the difference in complexity to the difference in energy between the states. 

\subsection{Bulk solutions}

\begin{figure}[htb]
\centering 
\includegraphics[width=0.45\textwidth]{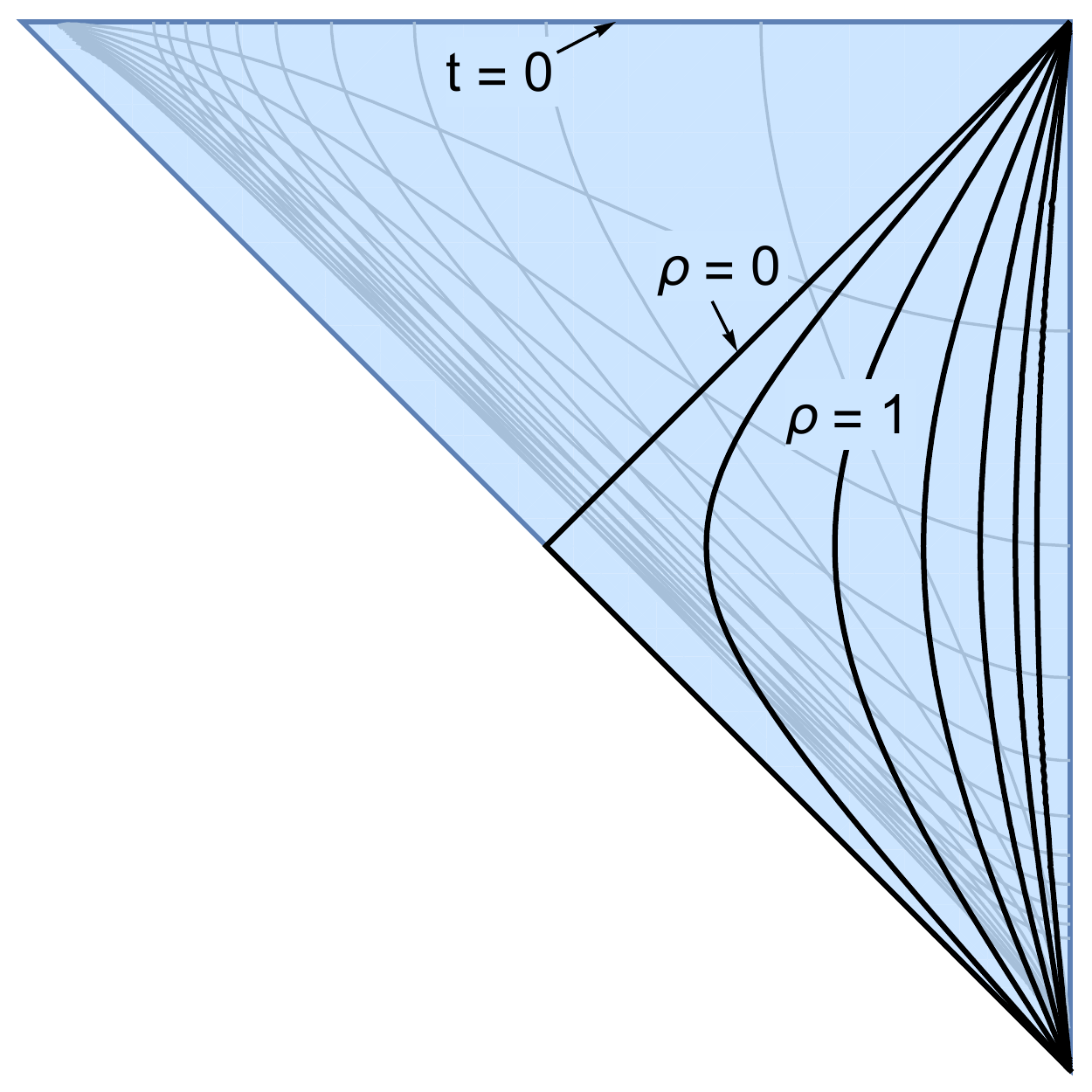}
\caption{The region of AdS covered by the de Sitter slicing \eqref{dsads}. The coordinates cover the right triangle, where surfaces of constant $\rho$ are plotted. The lower diagonal is the Poincar\'e horizon $z \to \infty, t \to - \infty$; the shaded region is the past half of the Poincar\'e patch. The de Sitter coordinates have an additional horizon at $\rho=0$. The region beyond this can be described by an FRW geometry of the form \eqref{frwads}. In pure AdS the horizontal line at the top is just $t=0$ in Poincar\'e coordinates. In the $\text{de Sitter} \times S^1$ examples we consider later, this is a ``singularity'' where the identification becomes null.}
\label{penrose}
\end{figure}

We study the field theory holographically by considering asymptotically anti-de Sitter spacetimes with a de Sitter boundary. The simplest example is a pure conformal field theory, which is simply dual to AdS$_d$ in the de Sitter slicing,
\begin{equation} \label{dsads}
ds^2 = \ell^2 (d \rho^2 + \frac{\sinh^2 \rho}{\eta^2} (-d \eta^2 + d\vec{x}^2)). 
\end{equation}
This is related to AdS in Poincar\'e coordinates \eqref{adsp} by 
\begin{equation} \label{fct}
z = -\frac{\eta}{\sinh \rho}, \quad t = \eta \coth \rho. 
\end{equation}
On the boundary, the $\eta$ and $t$ coordinates coincide, and the de Sitter and Poincar\'e coordinates are related simply by a conformal transformation that turns the $t=0$ slice in Poincar\'e coordinates into the future boundary of the de Sitter space, with the flat patch lying in the $t <0$ half of the Poincar\'e coordinates on the boundary. The de Sitter slices of constant $\rho$ in the coordinates \eqref{dsads} are surfaces of constant $t/z$ in the Poincar\'e coordinates, and there is a horizon at $\rho=0$, as shown in figure \ref{penrose}.

More generally, for asymptotically AdS solutions, the de Sitter symmetry implies that we can write the bulk geometry as 
\begin{equation}
ds^2 = \ell^2 (d \rho^2 + \frac{a^2(\rho)}{\eta^2} (-d \eta^2 +d\vec{x}^2) ),
\end{equation}
for some function $a(\rho)$ such that $a \to \sinh \rho$ as $\rho \to \infty$. 
%
%

There are two qualitatively different possibilities for the bulk geometry \cite{Maldacena:2012xp}: it could end at some $\rho$ with $a(\rho) >0$; this is referred to as a `gapped' phase, as it has the structure expected for a FT with an IR cutoff on de Sitter. Alternatively, the geometry could continue to a horizon, where $a(\rho)=0$. This is referred to as an `ungapped' phase. The pure AdS solution \eqref{dsads} is an example of an ungapped phase, but while pure AdS is smooth, generically for an ungapped phase we will have an FRW geometry beyond the horizon, 
\begin{equation} \label{frwads}
ds^2 = \ell^2 (- d \tau^2 + \frac{a^2(\tau)}{\eta^2} (d \eta^2 +d\vec{x}^2) ),
\end{equation}
with $a(\tau)$ rising from zero at the horizon to some maximum value, and then returning to zero at a ``big crunch'' singularity. 

We will focus our analysis on some states where bulk geometries are known analytically, which are obtained by considering the field theory on de Sitter$_{d-1} \times S^1$. A geometry with this boundary can be obtained by double analytic continuation of the Schwarzschild-AdS black hole. This gives the ``bubble of nothing'' spacetime \cite{Birmingham:2002st,Balasubramanian:2002am}
\begin{equation} \label{bubble}
ds^2 = f(r) d\chi^2 + f(r)^{-1} dr^2 + \frac{r^2}{\eta^2} (-d\eta^2 + d\vec{x}^2),
\end{equation}
with $f(r) =  1 + \frac{r^2}{\ell^2} - \frac{r_0^{d-2}}{r^{d-2}}$.  If we do a KK reduction over the $\chi$ circle, this provides an example of a geometry in the gapped phase. The reduced geometry is however singular, so it is simpler to analyse this case in the higher-dimensional geometry, where the geometry is smooth. The geometry ends at $r=r_+$, where $f(r_+) = 0$.  The circle direction closes off smoothly at this radius if we identify $\chi$ with period
\begin{equation} \label{bubblesize}
\Delta \chi = \frac{4\pi \ell^2 r_+}{ (dr_+^2 + (d-2)\ell^2)}. 
\end{equation}
This coordinate system is valid at all $r \geq r_+$, so there is no horizon in the bulk.  The quadratic relation between $\Delta \chi$ and $r_+$ implies that there is a maximum value of $\Delta \chi$ for which a bubble solution exists, $\Delta \chi_{\textnormal{max}} = 4 \pi \ell/ \sqrt{d (d-2)}$. For all smaller values of $\Delta \chi$, there are two bubble solutions, one with small $r_+$ and one with large $r_+$. The value of $r_0$ in terms of $r_+$ is $r_0^{d-2} = r_+^{d-2} (1+r_+^2/\ell^2)$. 

An alternative geometry with the same asymptotics is obtained by setting $r_0 = 0$ in the above solution. This solution is simply AdS with a periodic identification (it is the analytic continuation of thermal AdS, where the time circle becomes the $\chi$ circle in our spacetime). This could be rewritten as 
\begin{equation} \label{lads}
ds^2 = \cosh^2 \rho d\chi^2 + \ell^2 [d\rho^2 + \frac{\sinh^2 \rho}{\eta^2} (-d\eta^2 + d\vec{x}^2))] 
\end{equation}
by setting $r = \ell \sinh \rho$. In this case we can choose the period of $\chi$ freely. The fact that the geometry is locally AdS$_{d+1}$ makes the analysis of this case particularly straightforward. 

This is an example of an `ungapped' phase, with a horizon in the bulk spacetime. There is a coordinate singularity at $r= \rho =0$, but this is a horizon, and the geometry can be smoothly extended beyond it. This is clear if we consider the metric on surfaces of constant $\chi$: this is simply AdS in one lower dimension, in the de Sitter slicing \eqref{dsads}.  A good coordinate system which extends beyond this horizon is thus obtained by passing to Poincar\'e coordinates on this lower-dimensional AdS space, making the coordinate transformation \eqref{fct}. The metric is then 
\begin{equation} \label{ladsg}
ds^2 =  \frac{t^2}{z^2} d\chi^2 + \frac{\ell^2}{z^2} (- dt^2 + dz^2 + d\vec x^2 ) .
\end{equation}
We see that the identification along $\chi$ becomes null at $t=0$. As for a BTZ black hole, we cut off the spacetime at the ``singularity" where the circle identification becomes null, so the Penrose diagram for this spacetime is as shown in figure \ref{penrose}. In both solutions, it is the $\chi$ circle which is determining where we terminate the geometry, along a surface which would appear as a singularity from a dimensionally-reduced perspective. 

The boundary stress tensor for these solutions was calculated in \cite{Balasubramanian:2002am} for $d=4$. We work in a conformal frame where $\Delta \chi$ is the proper size of the $\chi$ circle, so the boundary metric is 
\begin{equation} 
ds^2_\partial = d\chi^2 + \frac{\ell^2}{\eta^2} (-d\eta^2 + d\vec x^2). 
\end{equation}
The stress tensor is (recalling that we work in units where $16 \pi G_N = 1$) 
\begin{equation} \label{bubblestress}
T^\chi_\chi = - \frac{3}{\ell^3} (r_0^2 + \ell^2/4), \quad T^\eta_\eta = T^{x_i}_{x_i} = \frac{1}{\ell^3} (r_0^2 + \ell^2/4). 
\end{equation}
The positive $T^\eta_\eta$ component corresponds to a negative energy density. For the values of $\Delta \chi$ where bubble solutions exist, we see that the bubbles have lower energy density than the locally AdS solution. The bubble with large $r_+$ is the lowest energy solution for a given $\Delta \chi$. This is analogous to the situation for a flat boundary with a circle direction, where the horizonless AdS soliton has a negative energy density, giving it lower energy than the identified Poincar\'e-AdS solution \cite{Horowitz:1998ha}. The difference in energy density between a bubble solution and the locally AdS solution is, for $d=4$, 
\begin{equation}
\Delta \rho = - \frac{r_0^2}{\ell^3}  = - \frac{r_+^2(1+r_+^2/\ell^2)}{\ell^3}. 
\end{equation}

We will compare the energy differences between these solutions to the complexity differences we calculate below. In the conformal frame adopted above, the energy of the state is simply
\begin{equation}
E = \frac{V_x \Delta \chi \ell^{d-2}}{|\eta|^{d-2}} \rho. 
\end{equation}
It is then convenient to write the complexity as in \eqref{cdens}, in terms of a complexity density $c$:  
\begin{equation}
\mathcal C  = \frac{V_x \Delta \chi \ell^{d-2}}{|\eta|^{d-2}} c. 
\end{equation}
The first factor is the proper volume of a given slice of the boundary in the conformal frame above, and $c$ is the complexity density, which carries the state-dependent information, and is a constant independent of the boundary coordinates. The derivative with respect to proper time gives
\begin{equation} \label{tcsymm}
\frac{d \mathcal C}{dt} = - \frac{\eta}{\ell} \frac{d \mathcal C}{d\eta} = (d-2) \frac{V_x \Delta \chi \ell^{d-3}}{|\eta|^{d-2}} c, 
\end{equation}
so \eqref{ctd} becomes a relation between the complexity density and the energy density,
\begin{equation} \label{dsb}
c \leq \frac{2 \ell}{(d-2) \pi \hbar} \rho.
\end{equation}
In the next sections, we test whether this bound can be satisfied and saturated in the different cases. 

\section{Holographic volume calculations}
\label{volume}

We now turn to the holographic calculation of the complexity for the spacetimes with de Sitter boundary introduced above. In this section, we calculate the complexity from the volume of the maximal volume spatial slice in the bulk, following the CV conjecture. 

For the case of pure AdS$_d$, we can use the coordinate transformation \eqref{fct} to identify the boundary slice of a cutoff boundary at some de Sitter time with a slice in Poincar\'e coordinates at fixed time, so the maximal volume surface is simply the constant-time surface in Poincar\'e coordinates. We take a boundary slice at $\eta = -\eta_0$ (so $\eta_0 >0$) and $\rho = \rho_0$ in the flat de Sitter coordinates. This  corresponds to $t =- t_0 = -\eta_0 \coth \rho_0$ and $z_0 = \eta_0 / \sinh \rho_0$ in Poincar\'e coordinates. The maximal volume slice is the surface of constant $t$ (this surface is described by  $\eta = -\eta_0 \tanh \rho/\tanh \rho_0$ in the de Sitter coordinates), whose volume is 
\begin{equation} \label{volads}
V = \int_{t=t_0} \sqrt{h} \mathop{dz} \mathop{d\vec x} = \ell^{d-1} V_x \int_{z_0}^{\infty} \frac{\mathop{dz}}{z^{d-1}} = \frac{\ell^{d-1} V_x}{(d-2) z_0^{d-2}} = \ell \frac{V_x}{(d-2) (H \eta_0)^{d-2}}, 
\end{equation}
where $H^{-1} = \ell \sinh \rho_0$ is the scale of the de Sitter space at the UV cutoff. This exhibits the expected UV divergence of the complexity. 

Turning to theories where the conformal symmetry is broken, we consider the explicit examples of asymptotically locally AdS$_{d+1}$ spaces with de Sitter$_{d-1} \times S^1$ boundaries. The simpler case is the ungapped solution, as the bulk is locally AdS. We want to consider the maximal volume slice with boundary at $\eta = -\eta_0$ at large $\rho$ in the metric \eqref{lads}. This coordinate system does not cover the whole maximal volume slice; it will enter the region inside the ``horizon'' at $\rho = 0$. Therefore to analyse this we use the coordinate transformation \eqref{fct} to pass to the Poincar\'e coordinates on the slices of constant $\chi$, so 
\begin{equation} \label{bhp}
ds^2 = \frac{t^2}{z^2} d\chi^2 + \frac{\ell^2}{z^2} (dz^2 -d t^2 + d\vec x^2).
\end{equation}
We are looking for a maximal volume surface ending at $t=t_0 = -\eta_0 $ as  $z \to 0$. This will now not lie at fixed $t$. Let us take $t = t(z)$. Then 
\begin{equation}
V = \ell^{d-1} \Delta \chi V_x \int_{z_0}^\infty \mathop{dz} \frac{|t| \sqrt{1-\dot{t}^2}}{z^{d}},
\end{equation}
where again $z_0 = \eta_0/\sinh \rho_0$. Let us redefine the variables in this integration by $z= \eta_0 \bar{z}$, $t = \eta_0 \bar{t}(\bar{z})$. Then 
\begin{equation}  
V = \frac{ \ell^{d-1} \Delta \chi V_x }{\eta_0^{d-2}} \int_{\bar z_0}^\infty \mathop{d\bar z} \frac{|\bar t| \sqrt{1-\dot{\bar t}^2}}{\bar z^{d}}. 
\end{equation}
We see the time dependence required by the general symmetry argument. The  remaining integral factor is a function only of the UV cutoff scale. In particular it will also be independent of the scale $\Delta \chi$, which enters here just as an overall multiplicative factor. Thus the complexity density in this case,
\begin{equation}  \label{mvol}
c  = \frac{8 (d-1)}{\pi} \int_{\bar z_0}^\infty d \bar z \frac{|\bar t| \sqrt{1-\dot{\bar t}^2}}{\bar z^{d}},
\end{equation}
is a pure numerical factor, depending only on the UV cutoff $\bar z_0 = 1/\sinh \rho_0$. Unlike the pure AdS case, however, this will have a finite contribution in addition to the UV divergent part.

We can solve the equations of motion arising from \eqref{mvol} with appropriate boundary conditions. We want $\bar t \to -1$ as $\bar z \to 0$. The minimal surface will approach this as $\bar t \approx - 1 - \frac{1}{2(d-1)} \bar z^2 + \mathcal{O}(z^4)$. There is a maximal volume slice behind the horizon; in the FRW patch coordinates, 
\begin{equation}
ds^2 = \ell^2 (\cos^2\rho\, d\chi^2 - d\rho^2 + \frac{\sin^2 \rho}{\eta^2} (d \eta^2 + d\vec x^2)),
\end{equation}
this is at $\cos \rho_* = \frac{1}{\sqrt{d}}$. This corresponds to $t = -\frac{1}{\sqrt{d}} z$ in the Poincar\'e coordinates. Requiring that the slice approaches this surface at large $z$, we find numerically that for $d=4$ 
\begin{equation} \label{cvung} 
c = \frac{24}{\pi} \left[ \frac{1}{3 z_0^3} + \frac{1}{9 z_0} -0.03384 \right]. 
\end{equation}
The divergent terms are determined in terms of the boundary geometry by the calculations of \cite{Carmi:2016wjl}.

It is interesting to note that in more general ungapped geometries, there is non-trivial time dependence encoded in the scale factor $a(\tau)$ in the FRW region beyond the horizon, but the complexity calculation is not hugely sensitive to this; as in this example, the maximal volume slice approaches a limiting surface where the scale factor is maximised, and the behaviour of the complexity will be mainly determined by this maximal value of the scale factor, and not its full time dependence. 

Let us now consider the bubble of nothing, which provides an example of a gapped geometry. The metric is 
\begin{equation}
ds^2 = f(r)\, d\chi^2 + \frac{dr^2}{f(r)} + \frac{r^2}{\eta^2} (-d\eta^2 + d\vec x^2).
\end{equation}
We can describe the bulk maximal volume surface by $\eta = \eta_0 e^{s(r)}$ for some function $s(r)$, with $s(r) \to 0$ as $r \to \infty$. This surface closes off at $r=r_+$ where the $S^1$ shrinks to zero size. The induced metric on the surface is 
\begin{equation}
ds^2 = f(r)\, d\chi^2 + \frac{dr^2}{f(r)} (1 - r^2 f(r) {s'}^2) + \frac{r^2}{\eta_0^2} e^{-2s}\, d\vec x^2.
\end{equation}
To analyse smoothness at $r=r_+$, it is convenient to introduce a radial coordinate $\bar r$ with 
\begin{equation}
\bar r = \int_{r_+}^r \frac{dr}{r \sqrt{f}}; 
\end{equation}
then the induced metric is 
\begin{equation}
ds^2 = f(r) d\chi^2 + r^2 d \bar r^2 \left( 1 - \left( \frac{ d s}{d \bar r} \right)^{2} \right) + \frac{r^2}{\eta_0^2} e^{-2s} d\vec x^2,
\end{equation}
and this will be smooth at $r=r_+$ for a suitable choice of period of $\chi$ if 
\begin{equation}
\frac{ds(r_+)}{d \bar r} = 0. 
\end{equation}

The volume of this surface is
\begin{equation} \label{gappedv}
\begin{split}
V &= \Delta \chi V_x \int_{r_+}^{r_{\textnormal{max}}} \mathop{dr} \sqrt{1 - f(r) \frac{r^2}{\eta^2} {\eta'}^2} \left(\frac{r}{\eta} \right)^{d-2}\\
&= \frac{\Delta \chi V_x}{\eta_0^{d-2}} \int_{r_+}^{r_{\textnormal{max}}} \mathop{dr} r^{d-2} \sqrt{1 - f(r) r^2 {s'}^2} e^{-(d-2) s} \\
&=  \frac{\Delta \chi V_x}{\eta_0^{d-2}} \int \mathop{d\bar r} r^{d-1}  \sqrt{f} \sqrt{1 - \left(\frac{ d s}{d \bar r}\right) ^{2}} e^{-(d-2) s},
\end{split}
\end{equation}
showing again that we get the time dependence required by symmetry. The integral is independent of $\eta_0$, but will now depend on $\Delta \chi$ through the dependence on $r_+$ in $f(r)$. Thus, the complexity density
\begin{equation}
c = \frac{8(d-1)}{\pi } \int \mathop{d\bar r} \left( \frac{r}{\ell} \right)^{d-1}  \sqrt{f} \sqrt{1 - \left(\frac{ d s}{d \bar r}\right) ^{2}} e^{-(d-2) s}
\end{equation}
is a function of $r_+$, and the result for the two bubbles for a given $\Delta \chi$ will be different. The UV divergent contributions are identical to the ungapped case; the $r_+$ dependence enters only in the finite term. Thus, the difference in complexity between different solutions with the same boundary is finite. 

We can formulate the problem of extremising the integral in \eqref{gappedv} by writing it as a first-order system, in terms of $s(r)$ and $q(r) =  \frac{ds}{d \bar r}$. The system to solve  is then
\begin{equation}
\frac{ds}{d r} = \frac{q}{r \sqrt{f}}, \quad \frac{dq}{dr} =\left[  \frac{(d-2)}{r \sqrt{f}} - q \left( \frac{(d-1)}{r} +  \frac{1}{2} \frac{f'}{f} \right) \right] (1-q^2)
\end{equation}
with the boundary conditions $q(r_+) = 0$, $s \to  0$ as $r \to \infty$.

 \begin{figure}
\centering 
\includegraphics[width=0.6\textwidth]{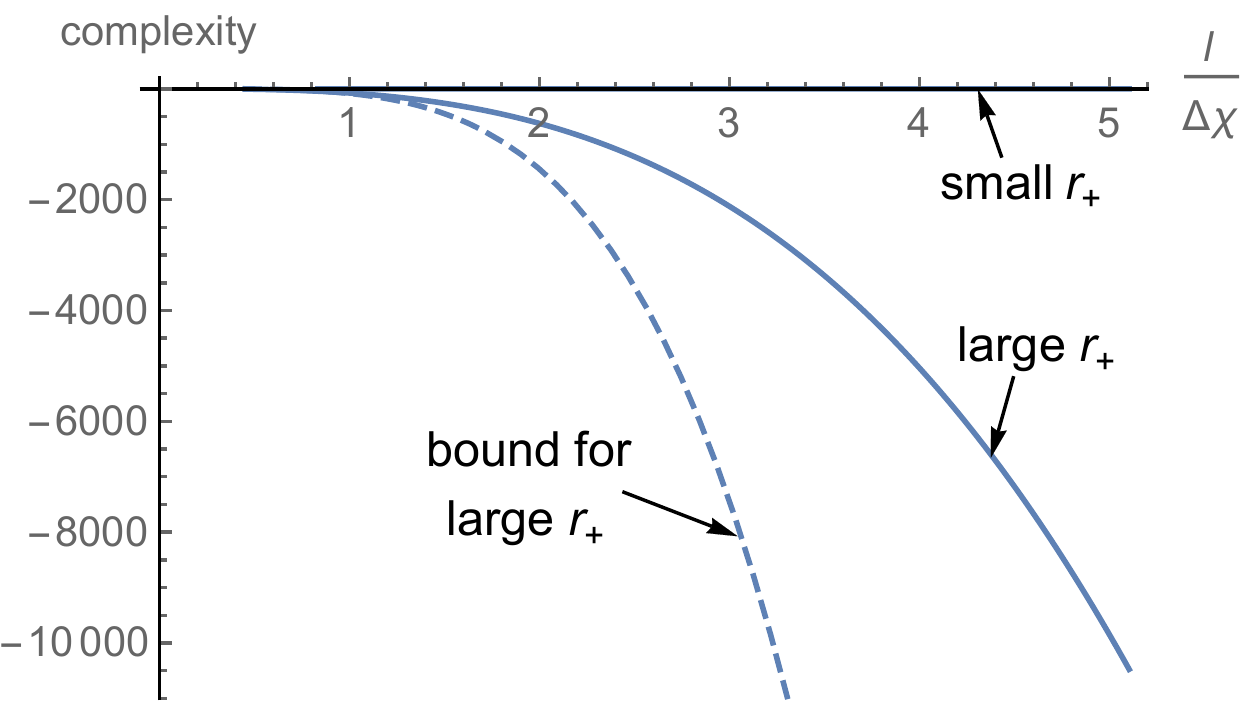}
\caption{The difference in complexity between the two bubble of nothing solutions and the  ungapped locally AdS solution, for $d=4$. We plot the complexity density $c$ as a function of $\ell/\Delta \chi$ (working with the inverse period makes the graph clearer). The upper branch is the small $r_+$ solutions, and the lower is the large $r_+$ solution. The dashed curve at the bottom is the bound on the difference in complexity from the difference in energy for the large $r_+$ solution. } \label{volplot}
\end{figure}

We plot numerical results for the two families of bubble solutions as a function of $1/\Delta \chi$ in figure \ref{volplot}. We see that the solution with larger $r_+$ has a smaller volume slice for a given value of $\Delta \chi$, as we would expect, since the surface of the bubble lies closer to the boundary. 

This leads to a smaller complexity, so lowering the energy of the state has lowered the complexity growth rate, as we would expect. The difference in complexity is however smaller than the difference in energy. The solid curve in figure \ref{volplot} plots the difference in the right-hand side of \eqref{dsb} between the ungapped solution and the large $r_+$ bubble. We see that the difference in complexity is smaller; hence if we assumed that the bubble solution satisfies the bound, the ungapped solution cannot saturate it. That is, the ungapped solution does not increase the complexity as fast as possible, given the energy difference from the large $r_+$ bubble, despite the presence of a horizon in the bulk.

This can be confirmed with an analytic argument. At large $r_+$, the function $f(r)$ is approximately 
\begin{equation} \label{fapp}
f(r) \approx \frac{r^2}{\ell^2} - \frac{r_+^d}{\ell^2 r^{d-2}} = \frac{r_+^2}{\ell^2} \tilde r ^2 \left( 1 - \frac{1}{\tilde r^d} \right), 
\end{equation}
where we have set $r = r_+ \tilde r$ in the second step to make the scaling clear. If we also define $s = \tilde s/r_+$, the volume integral becomes 
\begin{equation}
V = \frac{ \Delta \chi V_x}{\eta_0^{d-2}}  r_+^{d-1} \int_1^{\tilde r_{\textnormal{max}}} \mathop{d\tilde r} \tilde r^{d-2} \left[ 1 - \tilde r^4 \left( 1- \frac{1}{\tilde r^d} \right) \left( \frac{d \tilde s}{d \tilde r} \right)^2 \right]^{1/2} e^{-(d-2) \tilde s/r_+},
\end{equation}
where $\tilde r_{\textnormal{max}} = r_{\textnormal{max}} / r_+$. For large $r_+$, the exponential factor in the integral can be ignored. The function $\tilde s$ determined by extremising the integral then has no direct dependence on $r_+$. So dependence on $r_+$ enters only through the overall factor of $r_+^{d-1}$ and the cutoff $\tilde r_{\textnormal{max}} = r_{\textnormal{max}}/r_+$. The integral will have a UV divergence proportional to $\tilde r_{\textnormal{max}}^{d-1}$, which together with the $r_+^{d-1}$ prefactor gives the same $r_{\textnormal{max}}^{d-1}$ factor as before, so the UV divergent terms are independent of $r_+$, as they should be. The subleading divergences are absent in this approximation as our approximation for $f(r)$ neglects the curvature of the boundary. The finite contribution to the integral is independent of $r_+$, so the finite part of the volume will go as $r_+^{d-1}$ in this limit. 

Thus, the difference in complexity between the ungapped solution and the large bubble will scale as $r_+^{d-1}$. This grows more slowly with $r_+$ than the difference in energy, which goes as $r_+^d$, confirming the numerical results shown in figure \ref{volplot}, and extending them to general $d$.  A possible heuristic explanation of this behaviour of the complexity is that in a gapped theory, there is no structure on scales larger than $1/r_+$, so increasing $r_+$ is reducing the complexity by cutting out contributions at this scale, producing a reduction that scales like the volume in units of $1/r_+$.  It would be interesting to understand the difference from the black hole case in more detail. 

\begin{figure}
\centering 
\includegraphics[width=0.6\textwidth]{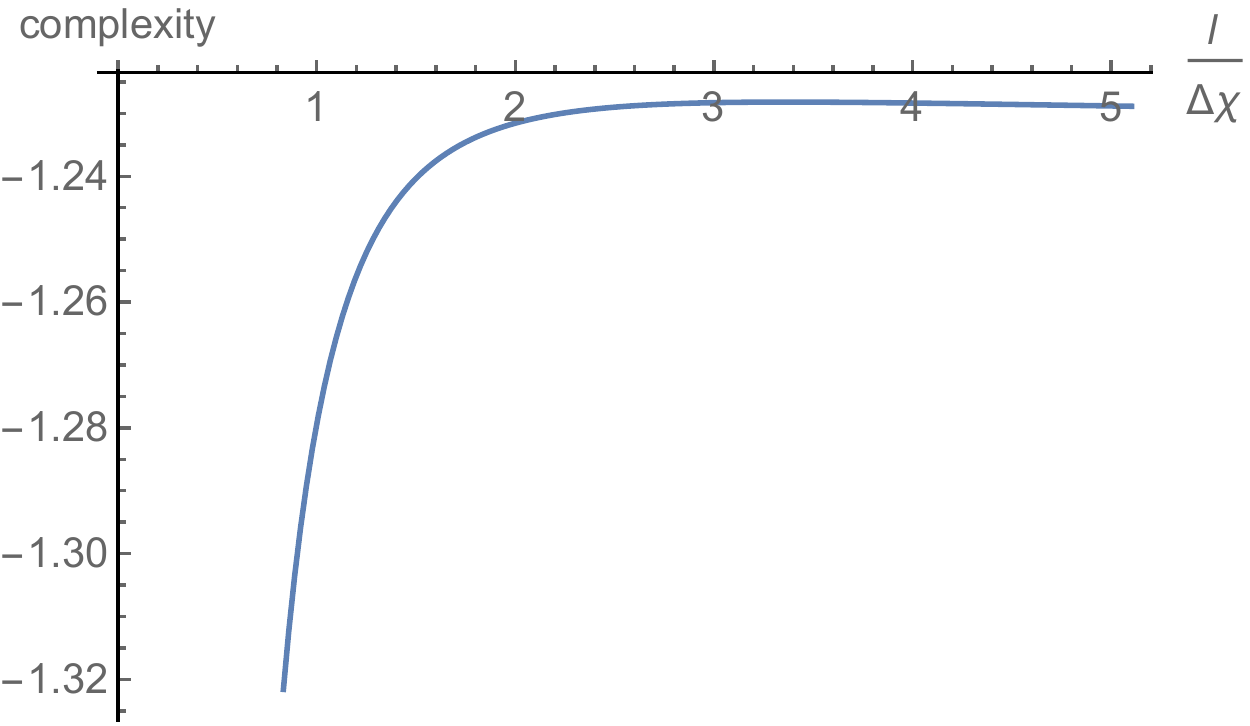}
\caption{The difference in complexity between the bubble of nothing solution with small $r_+$ and the ungapped locally AdS solution, for $d=4$. We plot the complexity density $c$, as a function of $\ell/\Delta \chi$. This is the same data as in figure \ref{volplot}, but plotted without the large $r_+$ data to zoom in on the difference in this case. We see that the difference does not approach zero for small $r_+$ (as we move to the right on the graph).} \label{volplot2}
\end{figure}

The volume of the slice in the bubble of small $r_+$ approaches a constant value for small $r_+$, that is, in the limit as $\Delta \chi \to \infty$. This is shown in figure \ref{volplot2}. However, there is a small finite difference between it and the ungapped solution in this limit, even though the difference in energy goes to zero. Geometrically this is unsurprising; the ungapped solution has an additional region of spacetime behind the horizon, so it is not surprising that the volume of the slice is larger. This is also consistent with expectations from the black hole examples that geometries with horizons will have more complexity for a given energy than those without. 

\section{Holographic action calculations}
\label{action}

We now turn to the calculation of the complexity using the CA conjecture, calculating  the action of the Wheeler-de Witt patch for these examples. Unlike in the black hole examples, where the action calculation agreed with the volume calculation up to coefficients, the action calculation for these examples gives qualitatively different results to the volume calculation. We will find that the action of the bubble solutions increases at large $r_+$, where the volume decreased, and exhibits a logarithmic divergence at small $r_+$. 

If we consider first pure AdS in de Sitter coordinates, the boundary slice at $\rho = \rho_0$, $\eta = \eta_0$ corresponds to $t = t_0 = -\eta_0 \coth \rho_0$ and $z_0 = \eta_0 / \sinh \rho_0$ in Poincar\'e coordinates, and the Wheeler-de Witt patch is just the one we calculated the action of before in \eqref{adsaf},\footnote{Note the difference in the power is because before we were discussing AdS$_{d+1}$, whereas here we've chosen to consider dS$_{d-1}$ slices in AdS$_d$.}
\begin{equation} 
S =  4  \frac{ \ell^{d-2}}{\epsilon^{d-2}} V_x \ln(d-2) =   4 \ln(d-2) \frac{V_x}{ (H \eta_0)^{d-2}},
\end{equation}
where, as before, $H^{-1} = \ell \sinh \rho_0$ is the scale of the de Sitter space at the UV cutoff. This has the same form as \eqref{volads}, up to the overall coefficient.

\begin{figure}[htbp]
\centering
\includegraphics[width=0.45\textwidth]{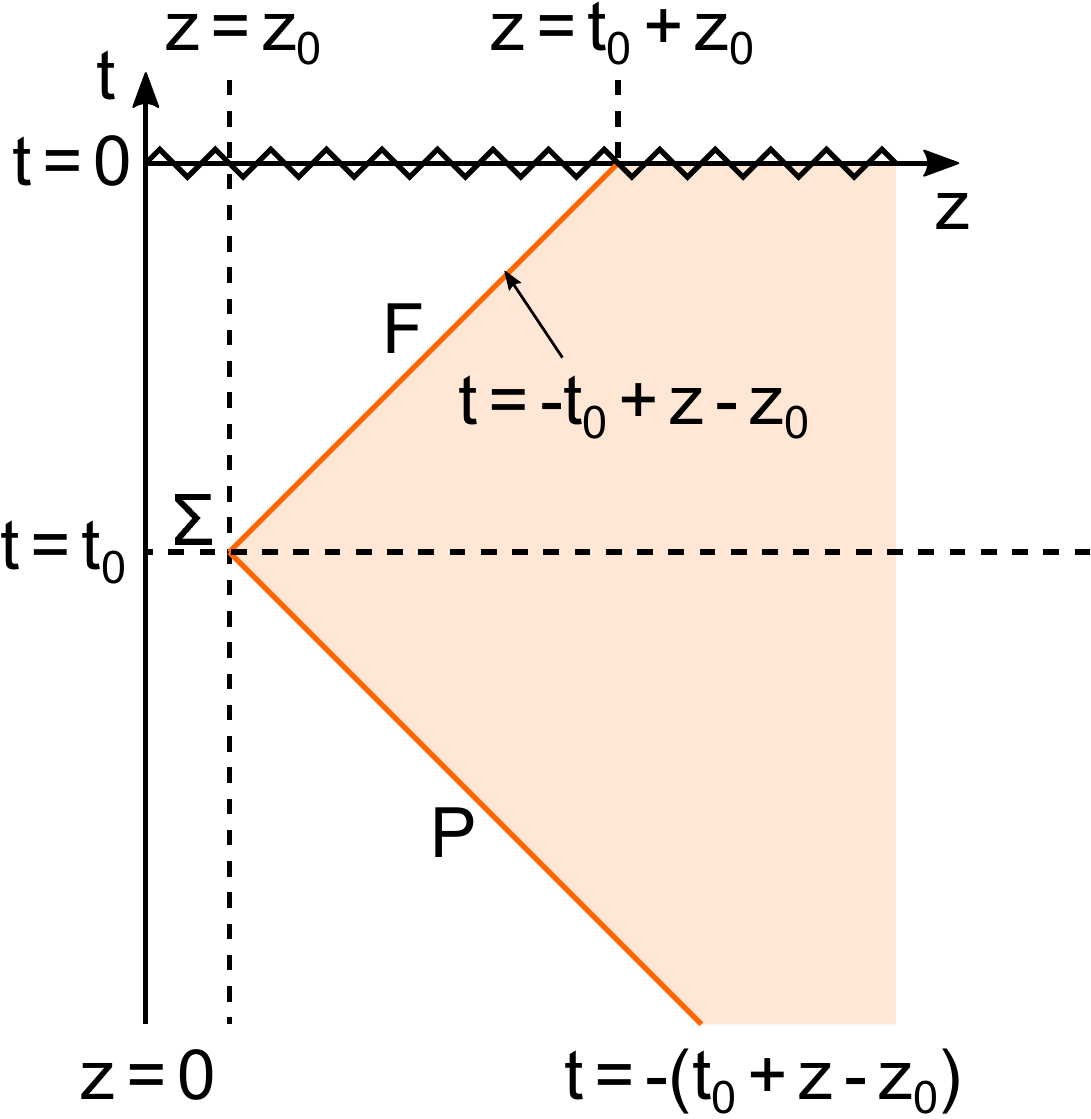}
\caption{The Wheeler-de Witt patch for the ungapped solution in Poincar\'e-like coordinates}
\label{BH WdW in Poincare}
\end{figure}

Let us turn now to the solutions with de Sitter$_{d-1} \times S^1$ boundaries. As before, the simple case is the ungapped solution, which is locally AdS$_{d+1}$. To calculate the action of the Wheeler-de Witt patch, it is again convenient to use the Poincar\'e coordinates, where the metric is 
\begin{equation} 
ds^2 = \frac{t^2}{z^2} d\chi^2 + \frac{\ell^2}{z^2} (dz^2 -d t^2 + d\vec x^2).
\end{equation}
We want the action of the Wheeler-de Witt patch for the boundary slice at  $\rho = \rho_0$, $\eta = -\eta_0$, which corresponds to $t = - t_0 = -\eta_0 \coth \rho_0$ and $z_0 = \eta_0 / \sinh \rho_0$. Near the boundary the Wheeler-de Witt patch looks just like in the case of pure AdS in Poincar\'e coordinates, with null surfaces at $dz = \pm dt$, but it is also cut off at $t=0$ as seen in the conformal diagram of figure \ref{penrose} and in figure \ref{BH WdW in Poincare}.
 
The volume integral is 
\begin{equation}
S_{\textnormal{Vol}} = - 2 d \ell^{d-2} V_x \Delta \chi \left( \int_{z_0}^{t_0+z_0} \frac{\mathop{dz}}{z^{d+1}} \int_{-(t_0+ z-z_0)}^{-t_0 + z- z_0}  |t| dt  +  \int_{t_0 +z_0}^\infty \frac{\mathop{dz}}{z^{d+1}} \int_{-(t_0 + z-z_0)}^{0} |t| dt  \right).
\end{equation}
Taking the affine parametrization 
\begin{equation} 
\lambda = - \frac{\ell^2}{\alpha z} \mbox{ on }F, \quad \lambda = \frac{\ell^2}{\beta z} \mbox{ on }P, 
\end{equation}
as before, $k = \alpha z^2/\ell^2 (\partial_t + \partial_z)$, $\bar k = \beta z^2/\ell^2 (\partial_t - \partial_z)$.  The boundary corner term is thus
\begin{equation}
S_{\Sigma} = - 2 \frac{\ell^{d-2} V_x \Delta \chi t_0}{z_0^{d-1}} \ln(\alpha \beta z_0^2/\ell^2).  
\end{equation}
There is a contribution from the spacelike surface along $t=0$, from $z  = t_0 + z_0$ to $z = \infty$,
\begin{equation}
S_{\textnormal{sing}} = -2 \int_{t_0+z_0}^\infty \mathop{dz} \sqrt{h} K =2 \ell^{d-2} V_x \Delta \chi \int_{z_0+t_0}^\infty \frac{\mathop{dz}}{z^{d-1}} .   
\end{equation}
Along the null boundaries, we include the additional contribution \eqref{acorr}. The metric on $F$ has 
\begin{equation}
\sqrt{\gamma} = \frac{ \ell^{d-2}}{z^{d-1}} |t| = \frac{\ell^{d-2}}{z^{d-1}}(t_0 - (z-z_0)), 
\end{equation}
so the expansion is
\begin{equation}
\Theta_F = \frac{1}{\sqrt{\gamma}} \frac{\partial \sqrt{\gamma}}{\partial \lambda} = - \frac{1}{\sqrt{\gamma}} \alpha  \frac{z^2}{\ell^2}  \frac{\partial \sqrt{\gamma}}{\partial z} = \frac{1}{\sqrt{\gamma}} \alpha  \frac{z^2}{\ell^2} \ell^{d-2} \left( (d-1) \frac{(t_0+z_0)}{z^{d}} - (d-2) \frac{1}{z^{d-1}} \right), 
\end{equation}
and hence the surface term is
\begin{equation}
S_{\textnormal{F}}  =  2 \ell^{d-2} V_x \Delta \chi  \int_{z_0}^{t_0+z_0}  \left( (d-1) \frac{(t_0+z_0)}{z^{d}} - (d-2) \frac{1}{z^{d-1}} \right) \ln (\ell \Theta_F) \mathop{dz}, 
\end{equation}
whereas on the past surface the expansion is
\begin{equation}
\Theta_P = \frac{1}{\sqrt{\gamma}} \alpha  \frac{z^2}{\ell^2} \ell^{d-2} \left( (d-1) \frac{(t_0-z_0)}{z^{d}} + (d-2) \frac{1}{z^{d-1}} \right) , 
\end{equation}
and the integral runs to infinity, i.e. 
\begin{equation}
S_{\textnormal{P}}  =  2 (d-2) \ell^{d-2} V_x \Delta \chi  \int_{z_0}^{\infty} \left( (d-1) \frac{(t_0-z_0)}{z^{d}} + (d-2) \frac{1}{z^{d-1}} \right) \ln(\ell \Theta_P) \mathop{dz}. \nonumber   
\end{equation}
Putting it all together, and setting $z_0 = \frac{\eta_0}{\sinh \rho_0}$, $t_0 = z_0 \cosh \rho_0$, the action for the Wheeler-de Witt patch is 
\begin{equation}
S = 2 \frac{V_x \Delta \chi}{(H \eta_0)^{d-2}} {\mathcal I}(\rho_0),  
\end{equation}
where, as before $H^{-1} = \ell \sinh \rho_0$, and the integral function is 
\begin{eqnarray}
{\mathcal I} &=& \frac{1}{(d-2)(1+\cosh \rho_0)^{d-2}}  + \int_1^{1+\cosh \rho_0} ( (d-1) (1+ \cosh \rho_0) - (d-2) \bar z) \ln (\theta_F) \frac{\mathop{d\bar z}}{\bar z^d} \nonumber \\ &&+ \int_1^\infty ((d-1) (\cosh \rho_0 - 1) + (d-2) \bar z) \ln(\theta_P) \frac{\mathop{d\bar z}}{\bar z^d}.
\end{eqnarray}
Here we have introduced new coordinates $\bar t = t/z_0$ and $\bar z = z/z_0$, and 
\begin{equation}
\theta_F = \bar z \frac{((d-1) (\cosh \rho_0+1) - (d-2) \bar z)}{\cosh \rho_0 + 1 - \bar z}, \quad \theta_P = \bar z \frac{((d-1) (\cosh \rho_0 -1) + (d-2) \bar z)}{\cosh \rho_0 -1 +\bar z}.
\end{equation}
We see that, as for the volume calculation, the time dependence is as determined by the symmetry, and the period $\Delta \chi$ appears only as an overall factor.  

The complexity density is 
\begin{equation}
c = \frac{2}{\pi} \sinh^{d-2} \rho_0 \, {\mathcal I}(\rho_0), 
\end{equation}
depending only on the UV cutoff as in the volume calculation \eqref{cvung}. For $d=3$, 
\begin{equation}
{\mathcal I} = 2 \ln 2 \cosh \rho_0 + \frac{1 + 2 \ln (2 \cosh \rho_0)}{4 \cosh \rho_0} + {\mathcal O}((\cosh \rho_0)^{-2}) . 
\end{equation}
For $d=4$,
\begin{equation}
{\mathcal I} =  2 \ln 3 \cosh \rho_0 + \frac{1}{3 \cosh \rho_0} + \frac{3 + \ln 16}{27 \cosh^2 \rho_0} + {\mathcal O}((\cosh \rho_0)^{-3}) . 
\end{equation}

It is interesting to note that for more general solutions with some arbitrary function $a(\rho)$, the complexity density in the action calculation will depend non-trivially on $a(\rho)$, and not just on the value at the limiting surface, as in the volume calculation. The action of the Wheeler-de Witt patch is a more sensitive probe of the bulk geometry than the volume of the maximal slice. 

Consider now the action of the Wheeler-de Witt patch in the bubble of nothing solutions. The main complication is in the position of the null boundaries. In the bubble metric \eqref{bubble}, if we take a slice of the boundary at $\eta = - \eta_0$, the null boundaries are given by 
\begin{equation}
\ln | \eta_{F,P}/\eta_0|  = \mp \int_r^{r_{\textnormal{max}}} \frac{\mathop{dr'}}{r' \sqrt{f(r')}}, 
\end{equation}
where along the future boundary, $|\eta_F| < \eta_0$, while along the past boundary, $|\eta_P| > \eta_0$, with $|\eta_{F,P}| \to \eta_0$ as $r \to r_{\textnormal{max}}$. The volume integral is 
\begin{equation} \label{svol}
\begin{split}
S_{\textnormal{Vol}} &= - \frac{2d V_{\vec x} \Delta \chi}{\ell^2} \int_{r_+}^{r_{\textnormal{max}}} r^{d-1} \mathop{dr} \int_{\eta_P}^{\eta_F}  \frac{\mathop{d\eta}}{|\eta|^{d-1}}\\
&= - \frac{2d V_{\vec x} \Delta \chi}{\ell^2 (d-2)} \int_{r_+}^{r_{\textnormal{max}}} r^{d-1} \mathop{dr} \left( |\eta_F|^{-(d-2)} - |\eta_P|^{-(d-2)} \right).
\end{split}
\end{equation}
If we write\footnote{For $d=4$, this can be written in terms of an elliptic integral, \[F(r) = \frac{\ell}{r_+} F\left(i \sinh^{-1} \left(r/\sqrt{\ell^2 + r_+^2}\right)\middle| -1- \frac{\ell^2}{r_+^2}\right)\].}
\begin{equation}
F(r) =  \int_r^{r_{\textnormal{max}}} \frac{\mathop{dr'}}{r' \sqrt{f(r')}}, 
\end{equation}
we can write this as 
\begin{equation}
S_{\textnormal{Vol}} =  - \frac{4d V_{\vec x} \Delta \chi}{\ell^2 (d-2) \eta_0^{d-2}} \int_{r_+}^{r_{\textnormal{max}}} r^{d-1} \sinh((d-2) F(r)) \mathop{dr}.   
\end{equation}

The tangent to the null surface is 
\begin{equation}
k = \alpha ( - \frac{\eta}{r^2} \mathop{\partial_\eta} - \frac{\sqrt{f(r)}}{r} \mathop{\partial_r}) 
\end{equation}
on $F$ and 
\begin{equation}
\bar k = \beta ( - \frac{\eta}{r^2} \mathop{\partial_\eta} + \frac{\sqrt{f(r)}}{r} \mathop{\partial_r}) 
\end{equation}
on $P$. As before, $\alpha$, $\beta$ are some arbitrary positive constants. Note that $\eta <0$, so the first terms are positive: these are future-pointing tangent vectors. The corner term in the action is 
\begin{equation}
S_\Sigma = - 2 V_{\vec x} \Delta \chi \sqrt{f(r_{\textnormal{max}})} \frac{r_{\textnormal{max}}^{d-2}}{ \eta_0^{d-2}} \ln\left( \frac{\alpha \beta}{r_{\textnormal{max}}^2} \right). 
\end{equation}

The expansions are
\begin{equation}
\Theta_F = - \alpha \frac{\sqrt{f}}{r} \frac{1}{\sqrt{\gamma}} \frac{\partial \sqrt{\gamma}}{\partial r} = - \alpha  \frac{\sqrt{f}}{r} \left( \frac{1}{2} \frac{f'}{f} + \frac{(d-2)}{r} - \frac{(d-2)}{r \sqrt{f}} \right) 
\end{equation}
and
\begin{equation}
\Theta_P = \beta \frac{\sqrt{f}}{r} \frac{1}{\sqrt{\gamma}} \frac{\partial \sqrt{\gamma}}{\partial r} = \beta  \frac{\sqrt{f}}{r} \left( \frac{1}{2} \frac{f'}{f} + \frac{(d-2)}{r} + \frac{(d-2)}{r \sqrt{f}} \right) 
\end{equation}
so the surface integrals are
\begin{equation} \label{sf}
S_{\textnormal{F}} = \frac{2 V_{\vec x} \Delta \chi}{\eta_0^{d-2}}  \int_{r_+}^{r_{\textnormal{max}}} \sqrt{f} r^{d-2} e^{(d-2)F} \left( \frac{1}{2} \frac{f'}{f} + \frac{(d-2)}{r} - \frac{(d-2)}{r \sqrt{f}} \right) \ln |\ell \Theta_F| \mathop{dr} 
\end{equation}
and
\begin{equation} \label{sp}
S_{\textnormal{P}} = \frac{2 V_{\vec x} \Delta \chi}{\eta_0^{d-2}}  \int_{r_+}^{r_{\textnormal{max}}} \sqrt{f} r^{d-2} e^{-(d-2)F} \left( \frac{1}{2} \frac{f'}{f} + \frac{(d-2)}{r} - \frac{(d-2)}{r \sqrt{f}} \right) \ln |\ell \Theta_P| \mathop{dr} 
\end{equation}
So the total integral is 
\begin{eqnarray}
S &=& \frac{2 V_{\vec x} \Delta \chi}{\eta_0^{d-2}}  \left[  - \frac{2d}{\ell^2 (d-2)}  \int_{r_+}^{r_{\textnormal{max}}} r^{d-1} \sinh((d-2) F(r)) \mathop{dr} - \sqrt{f(r_{\textnormal{max}})} r_{\textnormal{max}}^{d-2} \ln\left( \frac{\alpha \beta}{r_{\textnormal{max}}^2} \right) \right. \nonumber
\\ &&+\int_{r_+}^{r_{\textnormal{max}}} \sqrt{f} r^{d-2} e^{(d-2)F} \left( \frac{1}{2} \frac{f'}{f} + \frac{(d-2)}{r} - \frac{(d-2)}{r \sqrt{f}} \right) \ln |\ell \Theta_F| \mathop{dr}  \nonumber
\\ && \left. + \int_{r_+}^{r_{\textnormal{max}}} \sqrt{f} r^{d-2} e^{-(d-2)F} \left( \frac{1}{2} \frac{f'}{f} + \frac{(d-2)}{r} - \frac{(d-2)}{r \sqrt{f}} \right) \ln |\ell \Theta_P| \mathop{dr} \right]. 
\end{eqnarray}

The complexity density is 
\begin{eqnarray} 
c &=& \frac{2}{\pi}  \left[  - \frac{2d}{(d-2)}  \int_{r_+}^{r_{\textnormal{max}}} \left( \frac{r}{\ell} \right)^{d-1} \sinh((d-2) F(r)) \frac{\mathop{dr}}{\ell}  - \sqrt{f(r_{\textnormal{max}})} \left( \frac{r_{\textnormal{max}}}{\ell} \right)^{d-2} \ln\left( \frac{\alpha \beta}{r_{\textnormal{max}}^2} \right) \right. \nonumber
\\ &&+\int_{r_+}^{r_{\textnormal{max}}} \sqrt{f} \left( \frac{r}{\ell} \right)^{d-2} e^{(d-2)F} \left( \frac{1}{2} \frac{f'}{f} + \frac{(d-2)}{r} - \frac{(d-2)}{r \sqrt{f}} \right) \ln |\ell \Theta_F| \mathop{dr}  \nonumber
\\ && \left. + \int_{r_+}^{r_{\textnormal{max}}} \sqrt{f} \left( \frac{r}{\ell} \right)^{d-2} e^{-(d-2)F} \left( \frac{1}{2} \frac{f'}{f} + \frac{(d-2)}{r} - \frac{(d-2)}{r \sqrt{f}} \right) \ln |\ell \Theta_P| \mathop{dr} \right]. 
\end{eqnarray}

\begin{figure}
\centering 
\includegraphics[width=0.6\textwidth]{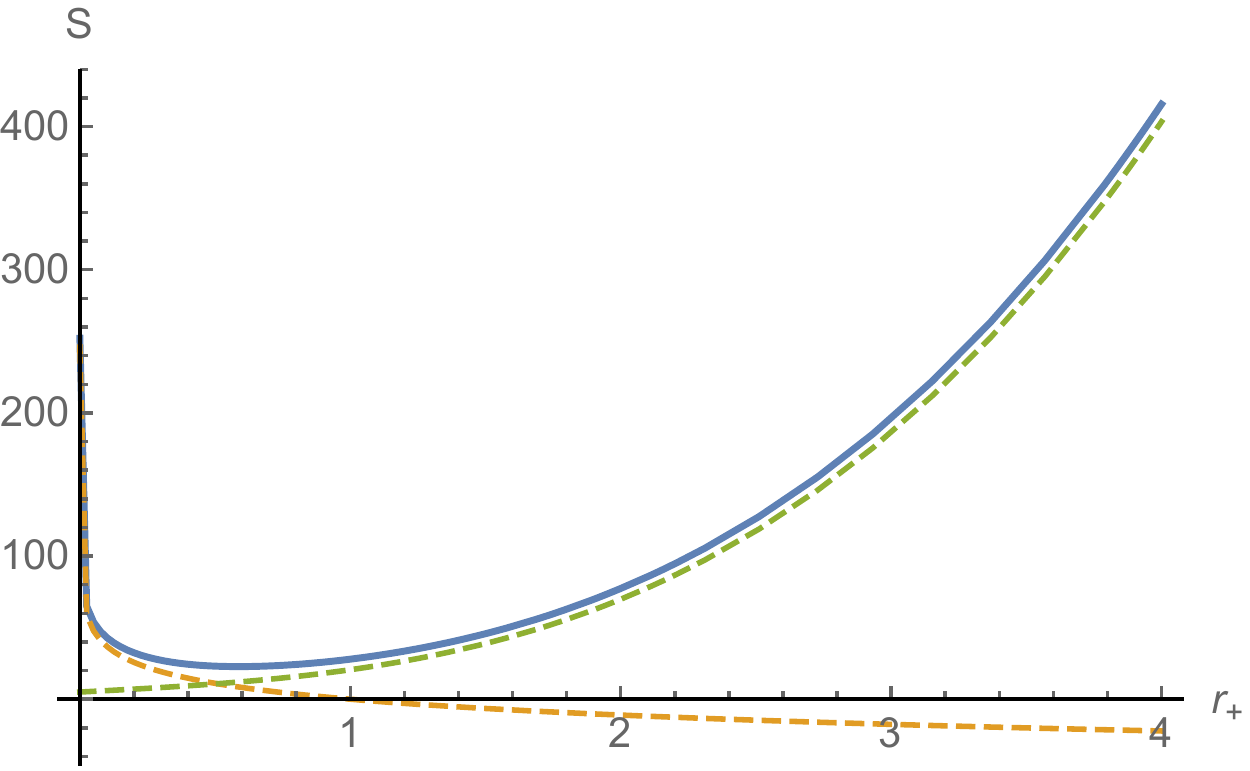}
\caption{The complexity density relative to the ungapped solution is plotted as a function of $r_+/\ell$ for the gapped solutions for $d=4$. The dotted curves indicate the asymptotics expected from the analytic discussion, with a fit to a function of the form $-a \ln r_+$ for small $r_+$ and $b r_+^3 + c r_+$ for large $r_+$. We see that the action of the bubbles is always greater than the ungapped solution, and increases for both large and small bubbles at small $\Delta \chi$. } \label{actionplot}
\end{figure}

As in the volume calculation, we want to compare the difference in complexity  between the two bubble solutions and the ungapped solution to the difference in energy. The action is calculated numerically and plotted in figure \ref{actionplot}, as a function of $r_+$. We see that the action {\it increases} relative to the ungapped solution, for both large and small bubbles. For large bubbles, the increase comes basically from the negative volume contribution; the smaller spacetime volume makes the negative contribution from \eqref{svol} less significant, increasing the action. The surface contributions (\ref{sf}, \ref{sp}) are numerically less important. Numerically, the bubble solutions always have a larger complexity than the ungapped solution. This is a surprising result; these solutions have lower energy, but larger complexity. For small bubbles, the increase comes from the additional surface terms we added on the null boundaries. 

As in the volume calculation, we can obtain the scaling for large $r_+$ from an analytic argument. Taking the approximation for $f(r)$ in \eqref{fapp}, we have $F(r) \approx \tilde F(\tilde r)/ r_+$, where 
\begin{equation} 
\tilde F = \int_{\tilde r}^{\tilde r_{\textnormal{max}}} \frac{\mathop{dr'}}{(r')^2 \sqrt{1 - (r')^{-d}}}.  
\end{equation}
The volume integral is 
\begin{equation} 
S_{\textnormal{Vol}} = r_+^{d-1} \int_1^{\tilde r_{\textnormal{max}}} \tilde r^{d-1} (d-2) \tilde F \mathop{d\tilde r} \sim r_+^{d-1} ( \tilde r_{\textnormal{max}}^{d-1} + \textnormal{finite}) \sim r_{\textnormal{max}}^{d-1} + O(r_+^{d-1}), 
\end{equation}
so the finite term is of order $r_+^{d-1}$. The corner contribution never makes a finite contribution. The contributions from $S_{\textnormal{F}}$ and $S_{\textnormal{P}}$ are a little more subtle; the expansions $\Theta_{F,P} \sim 1/r_+$, so in an expansion in $r_+$, $S_{\textnormal{F,P}}$ have terms of order $r_+^{d-1} \ln r_+$, but these are total derivatives:
\begin{eqnarray} 
S_{\textnormal{F,P}} &=& r_+^{d-1} \ln r_+ \int \Theta_{F,P} \sqrt{\gamma} \mathop{d\lambda} \mathop{d^{d-2}x} + O(r_+^{d-1})  \\
&=&  r_+^{d-1} \ln r_+ \int \partial_\lambda \sqrt{\gamma} \mathop{d\lambda} \mathop{d^{d-2}x} + O(r_+^{d-1}) \nonumber  \\ 
&=& r_{\textnormal{max}}^{d-1} \ln r_+ + O(r_+^{d-1}). \nonumber 
\end{eqnarray}
Crucially, there is no finite term in evaluating the total derivative integral, as $\sqrt{\gamma} = 0$ on the bubble. The divergent term combines with a $r_{\textnormal{max}}^{d-1} \ln \tilde r_{\textnormal{max}}$ divergence from the $r_+^{d-1}$ contribution to give the expected $r_{\textnormal{max}}^{d-1} \ln r_{\textnormal{max}}$ divergences in these integrals. So we have the leading UV divergences, which we know are independent of $r_+$ in general, and a finite term which comes just from the $r_+^{d-1}$ part. Thus the contributions from these surface integrals go as 
\begin{equation} 
S_{\textnormal{F,P}} \sim r_{\textnormal{max}}^{d-1}\ln r_{\textnormal{max}} +   r_{\textnormal{max}}^{d-1} +  O(r_+^{d-1}).
\end{equation}
The finite contributions to the action at large $r_+$ will scale like $r_+^{d-1}$, just as in the volume calculation. 

The crucial difference which we learn from the numerical analysis is that the sign is different. While in the volume calculation the finite contribution was reducing the volume as $r_+$ increased, here it is increasing the action. We can't fix this sign from the analytic scaling argument; it comes from the numerics. 

We can also use a similar argument to show that the action of the small bubbles has a divergence at small $r_+$. To carefully analyse the integration in the limit of small $r_+$, we will use two different approximations to $f(r)$: for $r \ll \ell$, we can write 
\begin{equation} 
f(r) \approx 1 - \frac{r_+^{d-2}}{r^{d-2}} = 1 - \frac{1}{\tilde r^{d-2}}, 
\end{equation}
where we again introduce a rescaled coordinate $r = r_+ \tilde r$, while for $r \gg r_+$, we can approximate
\begin{equation} 
f(r) \approx 1 + \frac{r^2}{\ell^2}. 
\end{equation}
For small $r_+$, these two approximations have an overlapping region of validity. In calculating the action, we can therefore divide the integration over $r$ into two regions, $r_+ \leq r < a$, and $a \leq r \leq r_{\textnormal{max}}$, for some $a$ such that $r_+ \ll a \ll \ell$, and we use the first approximation in the small $r$ regime and the second in the large $r$ regime. The overlap of the two approximations implies that the result will be independent of the particular value of $a$ where we choose to make the division. At the end of the calculation we want to take $r_+ \to 0$ at fixed, large $a/r_+$. The leading contribution from the second regime will be independent of $r_+$, so it is the contribution from the first regime that concerns us. 

For $r$ in the first region, $r_+ \leq r < a$, the approximation for $f(r)$ implies
\begin{equation} 
F(r) \approx \int_{\tilde r}^{a/r_+} \frac{\mathop{d\tilde r'}}{\tilde r' \sqrt{1 - 1/\tilde r^{' (d-2)}}} + \int_{a}^{r_{\textnormal{max}}} \frac{\mathop{dr'}}{r' \sqrt{1 + r^{'2}/\ell^2}} = - \ln{r_+} + \tilde F(\tilde r),
\end{equation}
where $\tilde F$ is finite as $r_+ \to 0$. Thus the volume integral from the first region is 
\begin{equation} 
S_{\textnormal{Vol}} \propto \int_{r_+}^a r^{d-1} \sinh((d-2) F) \mathop{dr} \approx r_+^2 \int_1^{a/r_+} \tilde r^{d-1} e^{(d-2) \tilde F} \mathop{d\tilde r},
\end{equation}
so the contribution from this integral vanishes as $r_+, a  \to 0$. The expansions in this region are
\begin{equation} 
\Theta_{F,P} = \frac{\tilde \theta_{F,P}}{r_+^2},  
\end{equation}
where $\tilde \theta_{F,P}$ is finite as $r_+ \to 0$.  In the integral on the past surface, the exponential in $F(r)$ makes the integrand small, so the contribution from the first region is finite:  
\begin{equation} 
S_{\textnormal{P}} \propto \int_{r_+}^a r^{d-1} e^{-(d-2) F} \Theta_P \ln |\ell \Theta_P| dr = r_+^{2d-5} \int_1^{a/r_+} \tilde r^{d-1} e^{-(d-2) \tilde F} \tilde \theta_P \ln | \ell \tilde \theta_P / r_+^2| d \tilde r. 
\end{equation}
the overall factor of $r_+$ ensures that this has no divergent contributions. However, in the integral on the future surface, the exponential in $F(r)$ cancels factors of $r_+$, giving 
\begin{equation} 
S_{\textnormal{F}} \propto \int_{r_+}^a r^{d-1} e^{(d-2) F} \Theta_F \ln |\ell \Theta_F| dr =  \int_1^{a/r_+} \tilde r^{d-1} e^{(d-2) \tilde F} \tilde \theta_F \ln | \ell \tilde \theta_F / r_+^2| d \tilde r. 
\end{equation}
There is a logarithmic divergence in $r_+$ here. The coefficient is a total derivative, as
\begin{equation} 
\Theta_{F} =  - \alpha \frac{\sqrt{f}}{r} \frac{1}{\sqrt{\gamma}} \partial_r \sqrt{\gamma} = - \frac{1}{r^{d-1} \eta^{d-2}}  \partial_r \sqrt{\gamma}, 
\end{equation}
so
\begin{eqnarray} 
S_{\textnormal{F}} &\propto& - 2 \ln r_+ \int_{r_+}^a \partial_r \sqrt \gamma \mathop{dr} + \ldots \\
&=& -2 \ln r_+ \sqrt{\gamma}(a) + \ldots \\ 
&=& -2 \ln r_+ \frac{1}{\eta_0^{d-2}} + \ldots, 
\end{eqnarray}
where in the first step we used the vanishing of the volume in the $\chi$ direction at $r=r_+$, and in the second step it is useful to note that the determinant of $\gamma$  is approximately constant in the region $r_+ \ll r \ll \ell$, so the result is independent of the particular value of $a$ chosen. Thus, the complexity for small $r_+$ grows like $- \ln r_+$. This is confirmed by the numerical results, which show a growth at small $r_+$. 

Thus, for small values of the period $\Delta \chi$, the CA calculation with our current prescription for the action gives that:
\begin{itemize}
\item For the ungapped solution, the complexity density is a constant, independent of $\Delta \chi$.
\item For the small bubble, the complexity density has a finite contribution on top of the ungapped result which grows as $-\ln \Delta \chi$ at small $\Delta \chi$. In this regime the energy density for the small bubble approaches the same value as for the ungapped solution. This divergence comes from the additional term on the future null surface we added to restore reparametrization invariance and eliminate undesirable log divergences in the UV. 
\item For the large bubble, the complexity density has a finite contribution on top of the ungapped result which grows as $1/\Delta \chi^{d-1}$ at small $\Delta \chi$. The energy density for the large bubble is less than that of the ungapped solution by a factor which grows as $1/\Delta \chi^{d}$. The action grows for larger bubbles because the volume integral makes a negative contribution to the action, and this dominates over the surface terms. 
\end{itemize}

These results are unexpected, and qualitatively different from what we obtained in the CV calculation. This might lead us to question whether the calculation we are applying is correct. The action we have considered is the one constructed in \cite{Lehner:2016vdi}, with our additional surface term. This is the minimal action whose variation vanishes for arbitrary variations of the metric holding the intrinsic geometry of the boundary fixed, and which is reparametrization invariant. However, we are free to add boundary terms to the action which are functions only of the intrinsic geometry of the boundary, obtaining alternative action whose variations also vanish. This includes reparametrization-invariant integrals along the null boundaries like the ones given in \eqref{acorr}. If we add such a term along the null boundaries, it will only add a time-independent correction to the action in the Schwarzschild-AdS case, so it would not modify the success of \cite{Lehner:2016vdi} in reproducing the expected behaviour \eqref{ctd}. But it could modify the value of the $r_+$ dependent part in our calculation. The challenge is to find an appropriate well-motivated correction. 

Note that although the ambiguity in the action we are considering here is the same one exploited in holographic renormalization, our case is different in that we are considering adding boundary terms on the boundary of the Wheeler-de Witt patch, which extends into the interior of the spacetime. Thus, these modifications can affect the finite, state-dependent part of the action. 

For our application to de Sitter in flat coordinates, the metric $\gamma_{ab}$ on the spatial slices of the null surface is flat, so it is challenging to find natural corrections which will change the answer. Integrals of the form \eqref{acorr} which involve the curvature of $\gamma$ will vanish in our case. If we take \eqref{acorr} with $f(\gamma)$ just a constant, this is a total derivative. We then have
\begin{equation} 
S_{\textnormal{N}} = f \int \Theta \sqrt{\gamma} \mathop{d\lambda} \mathop{d\chi} \mathop{d^{d-2}x} = f \int \partial_\lambda \sqrt{\gamma} \mathop{d\lambda} \mathop{d\chi} \mathop{d^{d-2}x} = f \int_{\Sigma} \sqrt{\gamma} \mathop{d\chi} \mathop{d^{d-2}x}, 
\end{equation}
so this term amounts to adding a counterterm at the corner. In the last step we used again that the volume element on the $\chi$ circle vanishes at the bubble. Adding such counterterms at the corner does not change the finite part of the action. The only non-trivial structure on the null surface in our case is the expansion. We can construct new reparametrization-invariant integrands by considering non-polynomial combinations of the expansion and its derivative with respect to $\lambda$: for example, we could add a term like 
\begin{equation} 
S_{\textnormal{N}} = \int \frac{\partial_\lambda \Theta}{\Theta} \sqrt{\gamma} \mathop{d\lambda} \mathop{d\chi} \mathop{d^{d-2}x}, 
\end{equation}
but such terms seem fairly contrived; it is difficult to see how such an action prescription would arise from a simple underlying principle.  Thus, it appears challenging to give a satisfactory prescription for the action which would give results for the complexity that are qualitatively similar to those from the volume calculation in these bubble solutions. 

\section{Discussion}
\label{disc}

We have considered the holographic calculation of the complexity for a field theory on de Sitter space in a de Sitter-invariant state. The holographic dual can have different bulk solutions satisfying these boundary conditions, corresponding to different de Sitter-invariant states in the boundary field theory. 

We find that the holographic complexity of these de Sitter-invariant states on a given spatial slice is a multiple of the proper volume of the slice. The de Sitter invariance fixes the multiplicative factor to be a state-dependent constant, independent of de Sitter time. Holographically, the states can have dual bulk solutions which have a horizon, corresponding to ungapped field theory states, or with no horizon, corresponding to gapped states. We have considered the particular case of field theory on de Sitter$_{d-1} \times S^1$, where explicit solutions of both kinds are known: the ungapped solution is locally AdS, and has a horizon in the bulk analogous to the one in the BTZ black hole. The gapped solutions are ``bubbles of nothing'' obtained by double analytic continuation from Schwarzschild-AdS. 

We found that in the CV calculation, the gapped solutions have lower complexity than the ungapped case, but the difference in complexity is smaller than the difference in energy, so the bound \eqref{ctd} on the growth of the complexity can be satisfied, but not saturated in both geometries. For the large bubbles, the difference in complexity scales as $r_+^{d-1}$, while the difference in energy scales as $r_+^d$. It would be interesting to understand why this bound is saturated for the field theory on flat space but not on these de Sitter spaces. 

In the CA calculation, we found a surprising sign difference: the action {\it grows} for both larger bubbles (like $r_+^{d-1}$) and for smaller bubbles (like $- \ln r_+$), so the complexity of the bubbles will  be larger than that for the ungapped solution. This is not what we expected to find, and suggests that the prescription for the action we have used should be modified. The log divergence for smaller bubbles came from the new term which we argued in our previous work should be introduced to make the action reparametrization-independent. It could be cured by removing this term, although that would leave the problem of reparametrization-dependence and a different UV divergence in the action calculations compared to the volume calculations. The growth of the action for large bubbles seems to come mainly from the volume term in the action. There is, in principle, freedom to modify the action prescription by adding boundary terms which depend just on the intrinsic geometry of the boundary of the Wheeler-de Witt patch. But it is challenging to find natural modifications which will make a difference in our case. We leave the problem of finding a resolution of this tension in the CA calculation for future work.

\section*{Acknowledgements}

AR is supported by an STFC studentship. SFR is supported in part by STFC under consolidated grant ST/L000407/1.

\bibliographystyle{JHEP}
\bibliography{complexity}

\end{document}